\documentclass[12pt]{article}
\usepackage{amsmath}
\usepackage{graphicx,psfrag,epsf}
\usepackage{enumerate}
\usepackage{natbib}
\usepackage{url} 
\usepackage{bbm}
\usepackage[caption=false]{subfig}
\usepackage{rotating}
\usepackage{xcolor}

\usepackage{verbatim}

\usepackage[linesnumbered,ruled]{algorithm2e}

\RequirePackage{amsmath, amssymb}
\newcommand{\C}{\mathcal{C}}
\newcommand{\R}{\mathbb{R}}
\newtheorem{lemma}{Lemma}
\newtheorem{corollary}{Corollary}

\newtheorem{proposition}{Proposition}
\newtheorem{theorem}{Theorem}

\newcommand\restrict[1]{\raisebox{-.5ex}{$|$}_{#1}}

\usepackage{amsmath}
\DeclareMathOperator*{\argmin}{argmin}   

\newtheorem{definition}{Definition}

\newcommand{\arr}{\xleftarrow{}}
\newcommand{\iid}{\underset{\text{iid}}{\sim}}

\newcommand{\SAR}{\Sigma_{{\mathsf{AR}}}}
\newcommand{\SRW}{\Sigma_{{\mathsf{RW}}}}

\usepackage{todonotes}
\newif\ifJASA
\JASAfalse 

\newif\ifnotblind
\notblindtrue 

\ifJASA
\else

\fi

\begin{document}

\def\spacingset#1{\renewcommand{\baselinestretch}%
{#1}\small\normalsize} \spacingset{1}


{
  \title{\bf Detecting Abrupt Changes in the Presence of Local Fluctuations and Autocorrelated Noise}
\ifnotblind
  \author{
  Gaetano Romano\\
    Department of Mathematics and Statistics,\\ Lancaster University, Lancaster, UK\\~\\
    Guillem Rigaill \\
    Université Paris-Saclay, CNRS, INRAE, Univ Evry, \\ Institute of Plant Sciences Paris-Saclay (IPS2), \\ 91405, Orsay, France
    \\~\\
    Vincent Runge \\
    Université Paris-Saclay, CNRS, Univ Evry,\\ Laboratoire de Mathématiques et Modélisation d'Evry\\91037, Evry, France \\~\\
    Paul Fearnhead\\
    Department of Mathematics and Statistics,\\ Lancaster University, Lancaster, UK}
\fi
  \maketitle
}


\begin{abstract}
Whilst there are a plethora of algorithms for detecting changes in mean in univariate time-series, almost all struggle in real applications where there is autocorrelated noise or where the mean fluctuates locally between the abrupt changes that one wishes to detect. In these cases, default implementations, which are often based on assumptions of a constant mean between changes and independent noise, can lead to substantial over-estimation of the number of changes. We propose a principled approach to detect such abrupt changes that models local fluctuations as a random walk process and autocorrelated noise via an AR(1) process. We then estimate the number and location of changepoints by minimising a penalised cost based on this model. We develop a novel and efficient dynamic programming algorithm, DeCAFS, that can solve this minimisation problem; despite the additional challenge of dependence across segments, due to the autocorrelated noise, which makes existing algorithms inapplicable. Theory and empirical results show that our approach has greater power at detecting abrupt changes than existing approaches. We apply our method to measuring gene expression levels in bacteria.

\end{abstract}

\noindent%
\vfill

\ifJASA
\spacingset{1.45} 
\fi


\section{Introduction}\label{sec:Introduction}

Detecting changes in data streams is  a ubiquitous challenge across many modern applications of statistics. It has been identified as one of the key open problems for modern analysis of large data \cite[]{Frontiers2013} and is important in such diverse areas as bioinformatics \cite[]{Olshen:2004,Futschik}, ion channels \cite[]{Hotz:2013}, climate records \cite[]{Reeves:2007}, oceonographic data \cite[]{Killick:2010} and finance \cite[]{Kim:2012}. The most common and important change detection problem is that of detecting changes in mean, and there have been a large number of different approaches to this problem that have been proposed \cite[e.g.][amongst many others]{Olshen:2004,Killick,Fryzlewicz:2014,Frick:2014,Maidstone,eichinger2018mosum,Fearnhead,fryzlewicz2018tail}. Almost all of these methods are based on modelling the data as having a constant mean between changes and the noise in the data being independent. Furthermore, all changepoint methods require specifying some threshold or penalty that affects the amount of evidence that there needs to be for a change before an additional changepoint is detected. In general the methods have default choices of these thresholds or penalties that have good theoretical properties under strong modelling assumptions.

\begin{figure}[!t]
    \label{fig:well-log}
    \centering
    \includegraphics[width = .9\linewidth]{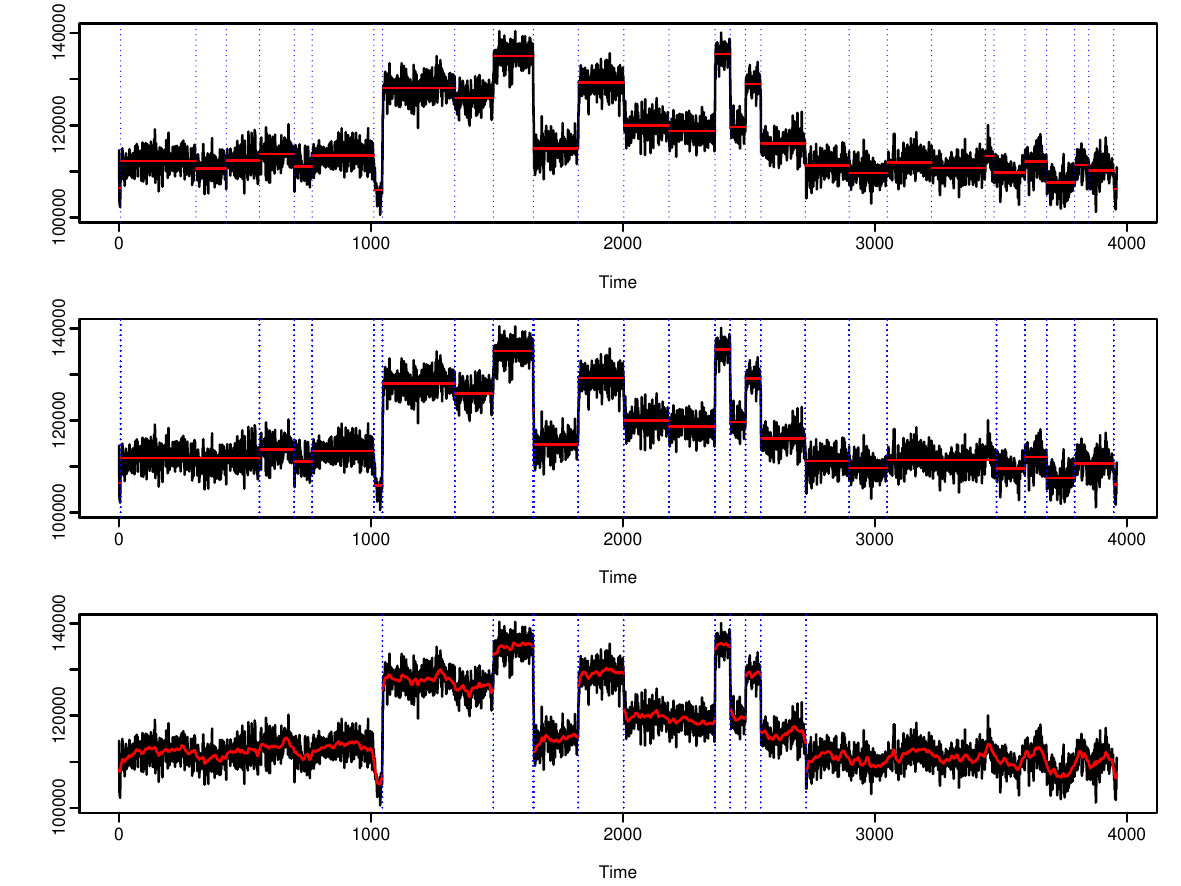}
    \caption{Segmentations of well-log data: wild binary segmentation using the strengthened Schwarz information criteria (top); segmentation under square error loss with penalty inflated to account for autocorrelation in measurement error (middle); optimal segmentation from DeCAFS with default penalty (bottom). Each plot shows the data (black line) the estimated mean (red line) and changepoint location (vertical blue dashed lines).
    }
\end{figure}

Whilst these methods perform well when analysing simulated data where the assumptions of the method hold, they can be less reliable in real applications, particularly if the default threshold or penalties are used. Reasons for this include the noise in the data being autocorrelated, or the underlying mean fluctuating slightly between the abrupt changes that one wishes to detect. To see this, consider change detection for the well-log data \cite[taken from][]{ruanaidh2012numerical,fearnhead2011efficient} shown in Figure \ref{fig:well-log}. This data comes from lowering a probe into a bore-hole, and taking measurements of the rock structure as the probe is lowered. The data we plot has had outliers removed. As the probe moves from one rock strata to another we expect to see an abrupt change in the signal from the measurements, and it is these changes that an analyst would wish to detect. Previous analyses of this data have shown that, marginally, the noise in the data is very well approximated by a Gaussian distribution; but by eye we can see local fluctuations in the data that suggest either autocorrelation in the measurement error, or structure in the mean between the abrupt changes.


The top plot shows an analysis of the well-log data that uses wild binary segmentation \cite[]{Fryzlewicz:2014} with the standard cusum test for a change in mean, and then estimates the number of changepoints based on a strengthened Schwarz information criteria. Both the cusum test and the strengthened Schwarz information criteria are based on modelling assumptions of a constant mean between changepoints and independent, identically-distributed (IID) Gaussian noise, and are known to consistently estimate the number and location of the changepoints if these assumptions are correct. However in this case we can see that it massively overfits the number of changepoints. Similar results are obtained for standard implementation of other algorithms for detecting changes in mean, see Figure \ref{fig:well-log-supp} in the Supplementary Material. 

\cite{lavielle2000least} and \cite{bardwell2019most} suggest that if we estimate changepoints by minimising the squared error loss of our fit with a penalty for each change, then we can correct for potential autocorrelation in the noise by inflating the penalty used for adding a changepoint. The middle plot of Figure \ref{fig:well-log} shows results for such an approach \cite[]{bardwell2019most}; this gives an improved result but it still noticeably overfits. 

By comparison, the method we propose models both autocorrelation in the noise and local fluctuations in the mean between changepoints -- and analysis of the data using default settings produces a much more reasonable segmentation of the data (see bottom plot of Figure \ref{fig:well-log}). This method is model-based, and assumes that the local fluctuations in the mean are realisations of a random walk and that the noise process is an AR(1) process. We then segment the data by minimising a penalised cost that is based on the  log-likelihood of our model together with a BIC penalty for adding a changepoint.

The key algorithmic challenge with our approach is minimising the penalised cost. In particular many existing dynamic programming approaches \cite[e.g.][]{jackson2005algorithm,Killick} do not work for our problem due to the dependence across segments caused by the autocorrelated noise. We introduce a novel extension of the functional pruned optimal partitioning algorithm of \cite{Maidstone}, and we call the resulting algorithm DeCAFS, for Detecting Changes in Autocorrelated and Fluctuating Signals. It is both computationally efficient (analysis of the approx 4000 data points in the well-log data taking a fraction of a second on a standard laptop) and guaranteed to find the best segmentation under our criteria.

Whilst we are unaware of any previous method that tries to model both autocorrelation and local fluctuations, \cite{chakar2017robust} introduced {AR1Seg} which aims to detect changes in mean in the presence of autocorrelation. Their approach is similar to ours if we remove the random walk component, as they aim to minimise a penalised cost where the cost is the negative of the log-likelihood under a model with an AR(1) noise process. However they were unable to minimise this penalised cost, and instead minimised an approximation that removes the dependence across segments. One consequence of using this approximation is that it often estimates two consecutive changes at each changepoint, and AR1Seg uses a further post-processing step to try and correct this. Moreover, our simulation results show that using the approximation leads to a loss of power, particularly when the autocorrelation in the noise is high.


The outline of the paper is as follows. In the next section we introduce our model-based approach and the associated penalised cost. In Section \ref{sec:algorithm-section} we present DeCAFS, a novel dynamic programming algorithm that can exactly minimise the penalised cost. To implement our method we need estimates of the model parameters, and we present a simple way of pre-processing the data to obtain these in Section \ref{sec:parameter-estimation}. We then look at the theoretical properties of the method. These justify the use of the BIC penalty, show that our method has more power at detecting changes when our model assumptions are correct than standard approaches, and also that we have some robustness to model error -- in that we can still consistently estimate the number and location of the changepoints in such cases by adapting the penalty for adding a changepoint. Sections \ref{sec:Simulation-Study} and \ref{sec:Real-Data} evaluate the new method on simulated and real data; and the paper ends with a discussion.

\ifnotblind
Code implementing the new algorithm is available in the R package \texttt{DeCAFS}. This package and full code from our simulation study is available at \url{github.com/gtromano/DeCAFS}.
\else 
Code implementing the new algorithm is available in the R package \texttt{DeCAFS}. This package and full code from our simulation study is available at \url{github.com/***}.

\fi
\section{Modelling and Detecting Abrupt Changes}
\label{sec:model-section}
\subsection{Model}

Let $y_{1:n} = (y_1, \dots, y_n) \in \mathbb{R}^n$ be a sequence of $n$ observations, and assume we wish to detect abrupt changes in the mean of this data in the presence of local fluctuations and autocorrelated noise. We take a model-based approach where  the signal vector is a realisation of a random walk process with abrupt changes, and we super-impose an AR(1) noise process. 

So for $t=1,\ldots,n$, 

\begin{equation}
    \label{eq:RWARmodel}
    y_t = \mu_t + \epsilon_t,
\end{equation}

where for $t=2,\ldots,n$

\begin{equation}
    \label{eq:RWmodel}
\mu_t = \mu_{t-1} + \eta_t + \delta_t, \quad \text{with} \ \eta_t \iid \mathcal{N}(0, \sigma_\eta^2), \ \delta_t \in \mathbb{R},
\end{equation}
and $\delta_t=0$ except at time points immediately after a  set of $m$ changepoints, $0 < \tau_1 < \cdots < \tau_m < n$. That is $\delta_t = 0$ unless $t = \tau_j+1$ for some $j$.
This model is unidentifiable at changepoints. If $\tau$ is a changepoint, then whilst the data is informative about $\mu_{\tau}$ and $\mu_{\tau-1}$, we have no further information about the specific value of $\delta_{\tau}$ relative to $\eta_{\tau}$. We thus take the convention that $\delta_\tau=\mu_{\tau}-\mu_{\tau-1}$ and $\eta_\tau=0$, which is consistent with maximising the likelihood for $\eta_{\tau}$.
The noise process, $\epsilon_t$ is a stationary AR(1) process with, for $t=2,\ldots,n$,
\begin{equation}
    \label{eq:ARmodel}
 \epsilon_t = \phi \epsilon_{t-1} + \nu_t \quad \text{with} \ \nu_t \iid \mathcal{N}(0, \sigma_\nu^2),
\end{equation}
for some autocorrelation parameter, $\phi$, such that $0\leq \phi < 1$; and $\epsilon_1\sim \mathcal{N}(0,\sigma_\nu^2/(1-\phi^2) ).$

\label{sec:edge-cases}
Special cases of our model occur when $\phi=0$ or when $\sigma^2_\eta = 0$.
When $\phi=0$ our noise process $\epsilon_t$ is then IID, and the model is equivalent to a random walk plus noise with abrupt changes. When $\sigma^2_\eta = 0$  we are detecting changes in mean with an AR(1) noise process, resulting in a formulation equivalent to the one of \cite{chakar2017robust}.

\subsection{Penalised Maximum Likelihood Approach}

In the following we will assume that $\phi$, $\sigma^2_\eta$ and $\sigma^2_\nu$ are known; we consider robust approaches to estimate these parameters from the data in Section \ref{S:EstimationParameter}. We can then write down a likelihood for our model as a function of $\mu_{1:n}$ and $\delta_{2:n}$. Writing $f(\cdot|\cdot)$ for a generic conditional density, 
we have that the likelihood is
\begin{eqnarray*}
\lefteqn{\mathcal{L}(y_{1:n};\mu_{1:n},\delta_{2:n})=\left(\prod_{t=2}^n f(\mu_t|\mu_{t-1},\delta_t)\right)f(y_1|\mu_1)\left(\prod_{t=2}^n f(y_t|y_{t-1},\mu_{t-1},\mu_t)\right)}\\
&\propto& \left(\prod_{t=2}^n \exp\left\{-\frac{(\mu_{t} - \mu_{t-1} - \delta_t) ^ 2 }{2\sigma_\eta^2}  \right\}  \right) \exp\left\{ -\frac{(y_1-\mu_1)^2}{2\sigma^2_{\nu}/(1-\phi^2)}\right\}\\
& &\times
\left(\prod_{t=2}^n \exp\left\{-\frac{((y_t - \mu_{t}) - \phi (y_{t-1} - \mu_{t-1}))^2 }{2\sigma_\nu^2} \right\} \right)\,.
\end{eqnarray*}
We have used the specific Gaussian densities of our model, and dropped multiplicative constants, to get the second expression.

If we knew the number of changepoints we could estimate their position by maximising this likelihood subject to the constraints on the number of non-zero entries of $\delta_{2:n}$. However, as we need to also estimate the number of changepoints we proceed by maximising a penalised version of the log of the likelihood where we introduce a penalty $\beta > 0$ for each changepoint -- this is a common approach to changepoint detection, see e.g. \cite{Maidstone}. It is customary to restate this as minimising a penalised cost, rather than maximising a penalised likelihood, where the cost is minus twice the log-likelihood. That is we estimate the number and location of the changepoints by solving the following minimisation problem: 

\begin{eqnarray}
\nonumber
      \lefteqn{ \mathcal{F}_n = \min_{\substack{\mu_{1:n}\\ \delta_{2:n}}} \Big\{
       (1-\phi^2) \gamma(y_1 - \mu_1)^2 \,\, + } \\
        && \sum_{t = 2}^n  \left[ \lambda (\mu_{t} - \mu_{t-1} - \delta_{t}) ^ 2 
         + \gamma \Big((y_t - \mu_t) - \phi (y_{t-1} - \mu_{t-1})\Big) ^ 2 + \beta \ \mathbbm{1}_{\delta_t \neq 0}\right] \Big\}, 
 \label{eq:penalised-cost}
\end{eqnarray}
where $\lambda = 1/\sigma_\eta^2$, $\gamma = 1/\sigma_\nu^2$, and $\mathbbm{1} \in \{0, 1 \}$ is an indicator function. For the special case of a constant mean between changepoints, corresponding to $\sigma_{\eta}^2=0$, we require $\mu_t = \mu_{t-1} + \delta_t \ \forall \ t = 2, \dots, n$ and simply drop the first term in the sum.

\subsection{Dynamic Programming Recursion}

We will use dynamic programming to minimise the penalised cost (\ref{eq:penalised-cost}). The challenge here is to deal with the dependence across changepoints due to the AR(1) noise process which means that some standard dynamic approaches for changepoint detection, such as optimal partitioning \cite[]{jackson2005algorithm} and PELT \cite[]{Killick}, cannot be used. To overcome this, as in \cite{Rigaill} or \cite{Maidstone}, we define the function $\mu \mapsto Q_t(\mu)$ to be the minimum penalised cost for data $y_{1:t}$ conditional on $\mu_t = \mu$, 

\begin{eqnarray*}
Q_t(\mu)& = & \min_{\substack{\mu_{1:t}\\ \delta_{2:t}, \mu_t=\mu}} \Big\{ 
        (1-\phi^2)\gamma(y_1-\mu_1)^2\,\,+  \\
       & & \sum_{i = 2}^t  \left[ \lambda (\mu_{i} - \mu_{i-1} - \delta_t) ^ 2 
         + \gamma \Big((y_i - \mu_i) - \phi (y_{i-1} - \mu_{i-1})\Big) ^ 2 + \beta \ \mathbbm{1}_{\delta_t \neq 0}\right] \Big\}. 
\end{eqnarray*}

So $\mathcal{F}_n=\min_{\mu \in \mathbb{R}} Q_n(\mu)$; and
the following proposition gives a recursion for $Q_t(\mu)$. 

\begin{proposition}
	\label{th:propositionNEW}
	
	The set of functions $\{\mu \mapsto Q_t(\mu)\,,\,t=1,\ldots,n\}$ satisfies
	
	$Q_1(\mu) = (1-\phi^2)\gamma(y_1-\mu)^2$ and, for $t=2,\ldots,n$,
	\begin{equation}
	Q_{t}(\mu) = \min_{\substack{u \in \mathbb{R}}} \left\{ Q_{t-1}(u) + \min\{\lambda (\mu - u) ^ 2, \beta\} + \gamma \Big((y_t - \mu) - \phi (y_{t-1} - u)\Big)^2 \right\} \,.
	 \label{updaterule} 
	\end{equation}
	
\end{proposition}

 The intuition behind the recursion is that we first condition on $\mu_{t-1}=u$, with the term in braces being the minimum penalised cost for $y_{1:t}$ given $u$ and $\mu_t=\mu$, and then minimise over $u$. The cost in braces is the sum of three terms: (i) the minimum penalised cost for $y_{1:t-1}$ given $u$; (ii) the cost for the change in mean from $u$ to $\mu$; and (iii) the cost of fitting data point $y_t$ with $\mu_t$. The cost for the change in mean, (ii), is just the minimum of the constant cost for adding a change and the quadratic cost for a change due to the random walk. The recursion applies to the special case of a constant mean between changepoints, where $\lambda=\infty$, if we replace $\min\{\lambda (\mu - u) ^ 2, \beta\}$ with its limit as $\lambda\rightarrow\infty$, which is $\beta\mathbbm{1}_{\mu \ne u}$.
\section{Computationally Efficient Algorithm}
\label{sec:algorithm-section}

\subsection{The DeCAFS Algorithm}

Algorithm \ref{alg:l2FPOP} gives pseudo code for solving the dynamic programming recursion introduced in  Proposition \ref{th:propositionNEW}. The key to implementing this algorithm is performing the calculations in line 5, and how this can be done efficiently will be described below. Throughout we give the algorithm for the case where there is a random walk component, i.e. $\lambda<\infty$, though it is trivial to adapt the algorithm to the $\lambda=\infty$ case.

As well as solving the recursion for $Q_t(\mu)$, Algorithm \ref{alg:l2FPOP} shows how we can also obtain the estimate of the mean, through a standard back-tracking step. The idea is that our estimate of $\mu_n$, $\hat{\mu}_n$, is just the value of $\mu$ that maximises $Q_n(\mu)$. We then loop backwards through the data, and our estimate of $\mu_{t}$ is the value that minimises the penalised cost for the data $y_{1:t}$ conditional on $\mu_{t+1}=\hat{\mu}_{t+1}$, which can be calculated as $B_t(\mu)$ in line 11. 

Finally, as we obtain the estimates of the mean, we can also directly obtain the estimated changepoint locations. It is straightforward to see, by examining the form of the penalised cost, that the optimal solution for $\delta_{2:n}$ has $\delta_{t+1}\neq 0$ (and hence $t$ is a changepoint) if and only if $\lambda(\hat\mu_{t+1}-\hat{\mu}_t)^2>\beta$.

\begin{algorithm}
	\label{alg:l2FPOP}
	\caption{DeCAFS}
	\KwData{${\bf y} = y_{1:n}$ a time series of length $n$}
	\KwIn{$\beta > 0$, $\lambda>0$, $\gamma>0$ and $0\leq \phi<1$.}
	\Begin(Initialisation){
		$Q_1(\mu) \arr (1-\phi^2)\gamma (y_1 - \mu) ^ 2$\\
	}
	\For{$t = 2$ to $n$}{
		$Q_{t}(\mu) \arr \underset{u}{\min} \left\{ Q_{t-1}(u) + \min\{\lambda (\mu - u) ^ 2, \beta\} + \gamma \Big((y_t - \mu) - \phi (y_{t-1} - u)\Big)^2 \right\}$\\
	}
	\Begin(Backtracking) {
		$\hat\mu_n\arr\argmin Q_n(\mu)$ \\
		$\hat\tau \arr n$ \\
		\For{$t = n-1$ to $1$}{
			$B_t(\mu) \arr Q_t(\mu)+\min\{\lambda(\mu-\hat\mu_{t+1})^2, \beta\} + \gamma\Big((y_{t+1}-\hat\mu_{t+1})-\phi(y_t-\mu)\Big)^2$\\
			$\hat\mu_t \arr \argmin B_t(\mu)$\\
			\If{$(\hat\mu_t  - \hat\mu_{t+1})^2 > \beta / \lambda$}{$\hat\tau \arr (t,\hat\tau)$}
		}
	}
	Return $\hat\mu_{1:n}$, $\hat\tau$
\end{algorithm}

\subsection{The Infimal Convolution} 

The main challenge with Algorithm \ref{alg:l2FPOP} is implementing line 5. Firstly this needs a compact way of characterising $Q_t(\mu)$. This is possible as $Q_1(\mu)$ is a quadratic function; and the recursion maps piecewise quadratic functions to piecewise quadratic functions. Hence $Q_t(\mu)$ will be piecewise quadratic and can be defined by storing a partition of the real-line together with the coefficients of the quadratics for each interval in this partition. 

Next we can simplify line 5 of Algorithm \ref{alg:l2FPOP}. As written it involves minimising a two-dimensional function, in $(u,\mu) \in \mathbb{R}^2$, over the variable $u$. We can recast this operation into a one-dimensional problem by introducing the concept of an infimal convolution.

\begin{definition} 
	\label{def:infimal_convolution}
	Let $f$ be a real-valued function defined on $\mathbb{R}$ and $\omega$ a non-negative scalar. We define $\mathsf{INF}_{f,\infty}(\theta)=f(\theta)$ and for $\omega > 0$,
	\begin{equation}
	\label{eq:inf_conv}
	\mathsf{INF}_{f,\omega}(\theta) = \min_{u\in \mathbb{R}}\left(f(u) + {\omega}(u-\theta)^2 \right),
	\end{equation}
	as the infimal convolution of $f$ with a quadratic term.
\end{definition}

A good review of this transformation can be found in Chapter $12$ of the book of \cite{bauschke2011convex}; it is closely related to the proximal mapping of $f$, which corresponds to the special case $\omega=1/2$. 

A property of the convolution is its stability for quadratics: the infimal transformation of a quadratic is a quadratic. Indeed, one can easily prove that the quadratic $q:\mu \mapsto a \mu^2 + b \mu + c$ with $(a,b,c) \in \R^+\times\mathbb{R}^2$ is transformed into 
$$
\mathsf{INF}_{q,\omega}:\mu \mapsto \frac{a\omega }{a + \omega} \mu^2 + \frac{b\omega }{a + \omega} \mu + c - \frac{b^2}{4 (a + \omega)}\,.
$$
We can also show that $q$ and $\mathsf{INF}_{q,\omega}$ have the same minimum and argminimum. Moreover, $\mathsf{INF}_{q,\omega} \le q$, resulting in a flattening of the quadratics.

The following proposition presents a reformulation of the update-rule into a minimization involving infimal convolutions. The proof is in Appendix \ref{App:Cost_Computation}.

\begin{proposition}
	\label{th:propositionALGO}
	
	The set of functions $\{Q_t(\mu)\,,\, t=2,\ldots,n\}$ can be written as
	\[
	Q_{t}(\mu) = \min \left\{ Q^{=}_{t}(\mu), \ Q^{\neq}_{t}(\mu)  \right\}, 
	\]
	where 
	$$
\begin{array}{l}
  Q^{=}_{t}(\mu) = \mathsf{INF}_{\mathbb{Q}_{t-1},\gamma\phi+\lambda}(\mu) +  \frac{\gamma}{1-\phi}\Big(y_t-\phi y_{t-1} - (1-\phi)\mu\Big)^2\,,\\
    Q^{\neq}_{t}(\mu) = \mathsf{INF}_{\mathbb{Q}_{t-1},\gamma\phi}(\mu) +  \frac{\gamma}{1-\phi} \Big( y_t-\phi y_{t-1} - (1-\phi)\mu\Big)^2 +\beta\,,
\end{array}
$$
	and
	\[
	\mathbb{Q}_{t-1}(u)=Q_{t-1}(u)- \gamma\phi(1-\phi)\left(u- \frac{y_t-\phi y_{t-1}}{1-\phi} \right)^2.
	\]
\end{proposition}

\subsection{Fast Infimal Convolution Computation}

As noted above we can represent $Q_t$ by  $\mathsf{Q}_t = (q_t^1,...,q_t^s)$ where each $q_t^i$ is a quadratic defined on some interval $[d_i,d_{i+1}[$ with $d_1 = -\infty$ and  $d_{s+1} = +\infty$. It is this representation of $Q_t$ that we update at each time step. Some operations involved in solving the recursion, such as adding a quadratic to a piecewise quadratic, or calculating the pointwise minimum of two piecewise quadratics are easy to perform with a computational cost that is linear in the number of intervals \cite[see e.g.][]{Rigaill} . 
The following theorem shows that a fast update for  the infimal convolution of a piecewise quadratic is also possible.

\begin{theorem}
\label{theorem}
Let $\mathsf{Q}_t = (q_t^1,...,q_t^s)$ be the representation of the functional cost $Q_t$. For all $\omega \ge 0$, the representation returned by the infimal convolution $\mathsf{INF}_{Q_t,\omega}$ has the following order-preserving form:

$$\mathsf{INF}_{\mathsf{Q}_t,\omega} = (\mathsf{INF}{q_t^{u_1}},\mathsf{INF}{q_t^{u_2}},...,\mathsf{INF}{q_t^{u_{s^*-1}}},\mathsf{INF}{q_t^{u_{s^*}}})\,,$$
with $1 = u_1 < u_2 < ... < u_{s^*-1} < u_{s^*} = s$ and $s^* \le s$.
\end{theorem}

The proof of this theorem is given in a general setting in Appendix \ref{App:InfProof}. 

Algorithm \ref{alg:pruning} shows how we can now calculate
$\mathsf{INF}_{\mathsf{Q}_t,\omega}$ in a linear-in-piece $O(s)$ time complexity. In this algorithm we have input $q_*^i = \mathsf{INF}{q_{t}^i}$, where $q_t^i$ is the $i^{th}$ piece-wise quadratic from $Q_t$ with $i \in \{1,...,s\}$. 
Algorithm \ref{alg:pruning} computes the intervals, $\textsc{dom}^i_*$ such that $\{\textsc{dom}^{u_i}_*, i = 1,...,s^*\}$  is the partition of the real line for $\mathsf{INF}_{\mathsf{Q}_t,\omega}$, with $Q_*$ storing the associated quadratics for each interval in this partition.
In Algorithm  \ref{alg:pruning} we use the list-operator $Last(l)$ to designate the last element of the list $l$; $index\,Last(l)$,
 $delete\,Last(l)$ to get the associated index of the last element or to delete this element. 

\begin{algorithm}[H]
    \label{alg:pruning}
    \SetKwInOut{Input}{Input}
    \Input{List of ordered quadratics $(q_*^1, q_*^2, \dots, q_*^{s-1}, q_*^s)$}
    \Begin(Initialization: $Q_*$ means "Remaining quadratics" and LB "Left Bound"){$Q_* \arr (q_*^1)$; $LB \arr (-\infty)$ \\}
    \For{$i = 2$ to $s$}{
    $j \arr index\,Last(Q_*)$\\
    $\mu_i : q^i_*(\mu_i) - q^j_*(\mu_i) = 0$ with $q^i_*(\mu) < q^j_*(\mu)$ for $\mu > \mu_i$ close to $\mu_i$\\
    \While{$\mu_i <$ Last(LB)}{
    $delete \, Last(Q_*)$; $delete \, Last(LB)$\\
     $j \arr index\,Last(Q_*)$\\
    $\mu_i : q^i_*(\mu_i) - q^j_*(\mu_i) = 0$ with $q^i_*(\mu) < q^j_*(\mu)$ for $\mu > \mu_i$ close to $\mu_i$\\}
    $Q_* \arr (Q_*, q_*^i)$;     $LB \arr (LB, \mu_i)$ \\
    }
    $s^* = \# LB$ (the number of element in $LB$)\\
    \For{$i = 1$ to $s^*-1$}{
	$\textsc{dom}_*^i = ]LB(i), LB(i+1)]$}
	$\textsc{dom}_*^{s^*}= ]LB(s^*), +\infty[$\\
    Return $Q_*$ and $(\textsc{dom}^1_*,...,\textsc{dom}^{s^*}_*)$
    
\caption{ $\mathsf{INF}_{\mathsf{Q}_t,\omega}$ pruning}
\end{algorithm}

\section{Robust Parameter Estimation} \label{S:EstimationParameter}
\label{sec:parameter-estimation}

Our optimisation problem (\ref{eq:penalised-cost}) depends on three unknown parameters: $\sigma^2_{\eta}$, $\sigma^2_{\nu}$ and $\phi$. We estimate these parameters by fitting to robust estimates of the variance of the $k$-lag differenced  data, $z_t^k = y_{t+k}-y_t$, for $k \ge 1$. 

\begin{proposition}
\label{eq:estim-proposition}
With the model defined  by (\ref{eq:RWARmodel}) -- (\ref{eq:ARmodel}), 
$$
z_t^k \sim \mathcal{N}\Big(\sum_{i=t+1}^{t+k}\delta_i, k\sigma_{\eta}^2 + 2\frac{1-\phi^k}{1-\phi^2}\sigma_{\nu}^2  \Big)\,,\quad t = 1,\dots ,n-k.$$

\end{proposition}

Providing $k$ is small relative to the lengths of segments, the mean of $z_t^k$ will be zero for most $t$. This suggests that we can estimate the variance of $z_t^k$ using a robust estimator, such as the median absolute difference from the median, or MAD, estimator. Fix $K$, and let $v_k$ be the MAD estimator of the variance of $z_t^k$ for $k=1,\ldots,K$. We estimate the parameters by minimising the least square fit to these estimates,
$$\mathcal{S}_{\phi}(\sigma_{\eta}^2,\sigma_{\nu}^2) = \sum_{k=1}^K \Big(k\sigma_{\eta}^2+ 2\frac{1-\phi^k}{1-\phi^2}\sigma_{\nu}^2 - v_k\Big)^2\,.$$
In practice we can minimise this criteria by using a grid of values for $\phi$ and then for each $\phi$ value analytically minimise with respect to $\sigma^2_\eta\geq0$ and $\sigma^2_\nu\geq0$. Obviously, if we are fitting a model without the random walk component we can set $\sigma^2_\eta=0$, or if we wish to have uncorrelated noise we set $\phi=0$. 

An empirical evaluation of this method for estimating the parameters is shown in the Supplementary material \ref{App:paramEstim}.
In our simulation study we use $K=15$, though similar results were obtained as we varied $K$.

\section{Theoretical Properties} \label{sec:Theory}

As is common with change-in-mean problems, we can reformulate our model as linear-regression. To do this it is helpful to introduce new variables, $\tilde{\eta}_{1:n}$, that give the cumulative effect of the random-walk fluctuations. To simplify exposition it is further helpful to define this process so it has an invertible covariance matrix. So we will let $\tilde{\eta}_1\sim \mathcal{N}(0,\sigma_\eta^2)$ and  $\tilde{\eta}_t=\tilde{\eta}_{t-1}+\eta_t$ for $t=2,\ldots,n$.
For a set of $m$ changepoints $\tau_{1:m}$, and defining $\tau_0=0$, we can introduce a $n\times (m+1)$ matrix $X_{\tau_{0:m}}$ where the $i$th column is a column of $\tau_{i-1}$ zeros followed by $n-\tau_{i-1}$ ones. 
Our model is then
\begin{equation} \label{eq:model}
y_{1:n}=X_{\tau_{0:m}}\Delta+\zeta_{1:n},
\end{equation}
where $\zeta_{1:n}$ is a vector of Gaussian random variables with
\[
\mbox{Var}(\zeta_{1:n})=\mbox{Var}(\epsilon_{1:n})+\mbox{Var}(\tilde{\eta}_{1:n}):=\SAR+\SRW
\]
the sum of the variance matrices for the AR component of the model, $\epsilon_{1:n}$, and the random walk component of the model, $\tilde{\eta}_{1:n}$; and $\Delta$ is a $(m+1)\times 1$ vector whose first entry is $\mu_1-\tilde{\eta}_1$ and whose $i$th entry is $\delta_{\tau_{i-1}+1}$ the change  at the $(i-1)$th changepoint. 

As shown in Appendix \ref{App:Theory}, the unpenalised version of the cost that we minimise, conditional on a specific set of changepoints, can be written as
\[
\mathcal{C}(\tau_{1:m})= \min_{\Delta,\tilde{\eta}_{1:n},\tilde{\eta}_1=0} \left[ 
(y_{1:n}-X_{\tau_{0:m}}\Delta-\tilde{\eta}_{1:n})^T \SAR^{-1} (y_{1:n}-X_{\tau_{0:m}}\Delta-\tilde{\eta}_{1:n})+\tilde{\eta}_{1:n}^T\SRW^{-1}\tilde{\eta}_{1:n}
\right],
\]
where $\tilde{\eta}_{1:n}$ is assumed to be a column vector. 
Thus the penalised cost (\ref{eq:penalised-cost}) is $\mathcal{F}_n=\min_{m,\tau_{1:m}}\left[ \mathcal{C}(\tau_{1:m})+m\beta \right]$. In the remainder of this section we will call $\mathcal{C}(\tau_{1:m})$ the cost, and $\mathcal{C}(\tau_{1:m})+m\beta$ the penalised cost.

Whilst our  cost is obtained by minimising over $\eta_{2:n}$, the following result shows that it is equal to the weighted residual sum of squares from fitting the linear model (\ref{eq:model}).
\begin{proposition} \label{prop:null}
The  cost for fitting a model with changepoints, $\tau_{1:m}$ is
\begin{equation} \label{eq:cost_prop}
\mathcal{C}(\tau_{1:m})=\min_{\Delta} (y_{1:n}-X_{\tau_{0:m}}\Delta)^T\left(\SAR+\SRW\right)^{-1}(y_{1:n}-X_{\tau_{0:m}}\Delta)
\end{equation} 
\end{proposition}
Let $\mathcal{C}_0$ denote the cost if we fit a model with no changepoints. The following corollary, which follows from standard arguments, gives the behaviour of the cost under a null model of no changepoints. This includes a bound on the impact of mis-specifying the covariance matrix, for example due to mis-estimating the parameters of the AR(1) or random walk components of the model, or if our model for the residuals is incorrect. 
\begin{corollary} \label{cor:null}
Assume that data is generated from model (\ref{eq:model}) with $m=0$ but with $\zeta_{1:n}$ a mean-zero Gaussian vector with $\mbox{Var}(\zeta_{1:n})=\Sigma$. Let  $\alpha^+_n$ be the largest eigenvalue of $(\SAR+\SRW)^{-1}\Sigma$. If $\Sigma=\SAR+\SRW$ then $\mathcal{C}_0-\mathcal{C}(\tau_{1:d})\sim \chi^2_d$. Otherwise,
for any $x$
\[
\Pr(\mathcal{C}_0-\mathcal{C}(\tau_{1:d})>x) \leq \Pr(  \chi^2_d>x/\alpha_n^+),
\]
Furthermore, if we estimate the number of changepoints using the penalised cost (\ref{eq:penalised-cost}) with penalty $\beta=C\alpha^+_n\log n$ for any $C>2$, then the estimated number of changepoints, $\hat{m}$, satisfies $\Pr(\hat{m}=0)\rightarrow 1$ as $n\rightarrow \infty$.
\end{corollary}

To gain insight into the behaviour of the procedure in the presence of changepoints, and how it differs from standard standard change-in-mean procedures, it is helpful to consider the reduction in cost if we add a single changepoint.
\begin{proposition} \label{prop:alt}
Given a fixed changepoint location $\tau_1$:
\begin{itemize}
    \item[(i)] The reduction in  cost for adding a single changepoint at $\tau_1$ can be written as
    \[
    \mathcal{C}_0-\mathcal{C}(\tau_1)= (v^Ty_{1:n})^2,
    \]
for some vector $v$ defined as
\[
v=\frac{1}{\sqrt{c_{\tau_1}-c_{0,\tau_1}^2/c_0}}\left\{(\SAR+\SRW)^{-1}u_{\tau_1}- \frac{c_{0,\tau_1}}{c_0}(\SAR+\SRW)^{-1}u_0 \right\},
\]
where $u_0$ is a column vector of $n$ ones, $u_{\tau_1}$ is a column vector of $\tau_1$ zero followed by $n-\tau_0$ ones, and
\[
c_0= u_0^T(\SAR+\SRW)^{-1}u_0, ~~~ c_{0,\tau_1}=u_0^T(\SAR+\SRW)^{-1}u_{\tau_1}, 
~~ c_{\tau_1}=u_{\tau_1}^T(\SAR+\SRW)^{-1}u_{\tau_1}.
\]
\item[(ii)] The vector $v$ in (i) satisfies $\sum_{i=1}^n v_i=0$ and $v^T(\SAR+\SRW)v=1$.
\item[(iii)] For any vector $w$ that satisfies $\sum_{i=1}^n w_i=0$ and $w^T(\SAR+\SRW)w=1$,
\[
\left(\sum_{i=\tau_1+1}^n w_i \right)^2\leq\left(\sum_{i=\tau_1+1}^n v_i \right)^2.
\]
\end{itemize}

\end{proposition}

The vector $v$ in part (i) of this proposition defines a projection of the data that is used to determine whether to add a changepoint at $\tau_1$. The properties in part (ii) mean that this projection is invariant to shifts of the data, and that the distribution of the reduction in cost if our model is correct and there are no changes will be $\chi^2_1$. The statistic $v^T y_{1:n}$ can be viewed as analogous to the cusum statistic \cite[]{hinkley1971inference} that is often used for a standard change-in-mean problem, and in fact if we set $\phi=0$ and $\sigma_\eta=0$ so as to remove the auto-regressive and random-walk aspects of the model, $|v^T y_{1:n}|$ is just the standard cusum statistic. The power of our method to detect a change at $\tau_1$ will be governed by the distribution of this projection applied to the data in the segments immediately before and after $\tau_1$. For a single changepoint where the mean changes by $\delta$ this distribution is a non-central chi-squared with 1 degree of freedom and non-centrality parameter $\delta^2(\sum_{i=\tau_1+1}^n v_i)^2$. Thus part (iii) shows that $v$ is the best linear projection, in terms of maximising the non-centrality parameter, over all projections that are invariant to shifts in the data and that are scaled so that the null distribution is $\chi^2_1$. 

\begin{figure}
	\centering
	\includegraphics[width=0.65\linewidth]{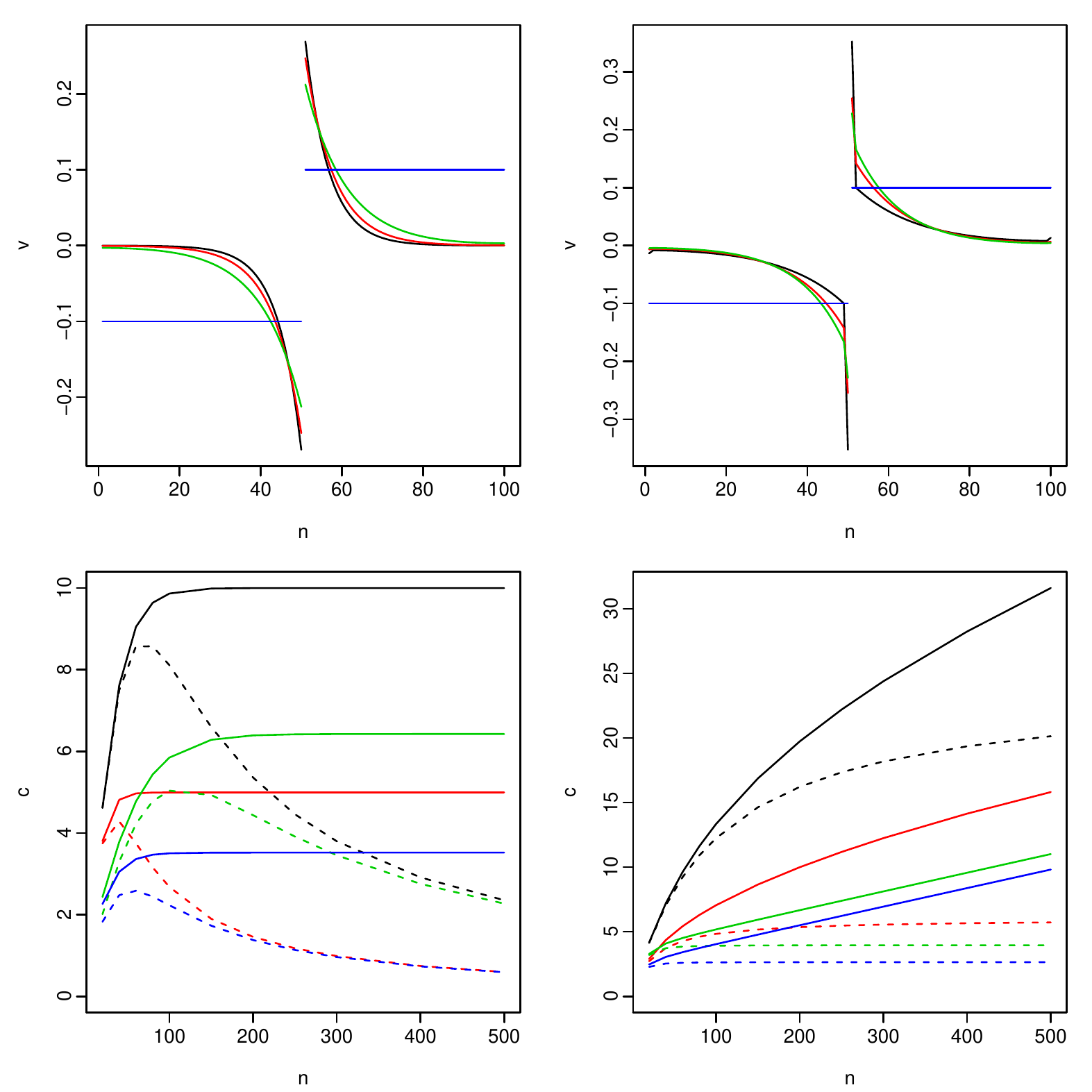}
	\caption{ Top row: projections of data $v$ for detecting a change in the middle of $n=100$ data-points. Random walk model (top-left) for varying $\sigma^2_\eta$ of 0.03 (black), 0.02 (red) and 0.01 (green); AR(1) plus random walk model (top-right) for $\sigma^2_\eta=0.01$ and varying $\phi$ of 0.4 (black), 0.2 (red) and 0.1 (green). In both plots the blue line shows the standard cusum projection. Bottom row: non-centrality parameter for a $\chi^2_1$ test of a change using the optimal projection (solid line) and the cusum projection (dashed line) for a change of size 1 in the middle of the data as we vary $n$. Out-fill asymptotics (bottom-left) where $(\sigma^2_\eta,\phi)$ is (0.0025,0)  (black), (0.01,0) (red), (0.0025,0.5) (green) and (0.01,0.5) (blue); In-fill asymptotics (bottom-right) where for $n=50$ $(\sigma^2_\eta,\phi)$ is (0.0025,0)  (black), (0.01,0) (red), (0.0025,0.5) (green) and (0.01,0.5) (blue).	} 
	\label{fig:projections}
\end{figure}

To gain insight into how the auto-regressive and random-walk parts of the model affect the information in the data about a change we have plotted different projections $v$ for different model scenarios in the top row of Figure \ref{fig:projections}. The top-left plot shows the projections if we have $\phi=0$ for different values of the random walk variance. The projection, naturally, places more weight to data near the putative changepoint, and the weight decays essentially geometrically as we move away from the putative changepoint. In the top-right plot we show the impact of increasing the autocorrelation of the AR(1) process, with the absolute value of the weight given to data points immediately before and after the putative change increasing with $\phi$. 

A key feature of the random walk model is that for any fixed $\sigma^2_\eta>0$ the amount of information about a change will be bounded as we increase the segment lengths either side of the change. This is shown in the bottom-left plot of Figure \ref{fig:projections} where we show the non-centrality parameter for detecting a change in the middle of the data as we vary $n$. For comparison we also show the non-centrality parameter of a test based on the cusum statistic (scaled so that it also has a $\chi^2_1$ distribution under the null of no change). We can see that ignoring local fluctuations in the mean, if they exist and come from a random walk model, by using the cusum statistic leads to a reduction of power as segment lengths increase. For comparison in the bottom right we show an equivalent comparison where we consider an infill asymptotic regime, so that as $n$ increases we let the random walk variance decay at a rate proportion to $1/n$ and we increase the lag-1 autocorrelation appropriately. In this case using the optimal projection gives a non-centrality parameter that increases with $n$, whereas the cusum statistic has power that can be shown to be bounded as we increase $n$.


We now turn to the property of our method at detecting multiple changes. Based on the above discussion, we will consider in-fill asymptotics  as $n\rightarrow\infty$.
\begin{itemize}
    \item[(C1)] Let $y_1,\ldots,y_n$ be generated as a finite sample from a Gaussian process on $[0,1]$; that is $y_i=z(i/n)$ where, for $t\in[0,1]$ $z(t)=\mu(t)+\zeta(t)$, $\mu(t)$ is a piecewise constant with $m^0$ changepoints at locations $r_1,\ldots,r_{m^0}$, and $\zeta(t)$ is a mean zero Gaussian process. For a given $n$ define the true changepoint locations as $\tau^0_i=\lfloor n r^0_i \rfloor$. The change in mean at each changepoint is fixed and non-zero.
    \item[(C2)] Assume there exists strictly positive constants $c_\eta$, $c_\nu$ and $c_\phi$, such that we implement DeCAFS with $\sigma^2_\eta=c_\eta/n$ and either
  (i) $\phi=0$ and $\sigma^2_\nu=c_\nu$; or
(ii) $\phi=\exp\{-c_\phi/n\}$ and $\sigma_{\nu}^2=c_\nu(1-\exp\{-2c_\phi/n\})$.
    \item[(C3)] There exists an $\alpha$ such that for any large enough $n$ if $\Sigma^0_n$ is the covariance of the noise in the data generating model (C1), and $\SAR^{(n)}+\SRW^{(n)}$ is the covariance assumed by DeCAFS in (C2) then the largest eigenvalue of $(\SAR^{(n)}+\SRW^{(n)})^{-1}\Sigma^0_n$ is less than $\alpha$.
\end{itemize}

The key condition here is (C3) which governs how accurate the model assumed by DeCAFS is to the true data generating procedure. Clearly if the model is correct then (C3) holds with $\alpha=1$. 
The following proposition gives upper bound on $\alpha$ in the the case where the covariance of the data generating model is that of a random walk plus AR(1) process, but with different parameter values to those assumed by DeCAFS in (C2).

\begin{proposition} \label{prop:C3}
Assume the noise process $\zeta(t)$ of the data generating process (C1) is equal to a random walk plus an AR(1) process. 
\begin{itemize}
\item[(i)] If $\mbox{Cov}(\zeta(t),\zeta(s))=c^0_\eta\min(t,s)$ for $t\neq s$ and $\mbox{Var}(\zeta(t))=c^0_\eta t + c_\nu$, and DeCAFS is implemented as in (C2)(i), then (C3) holds with $\alpha=\max\{c^0_\nu/c_\nu,c^0_\eta/c_\eta\}$.  
\item[(ii)] If $\mbox{Cov}(\zeta(t),\zeta(s))=c^0_\eta\min(t,s)+c_\nu^0\exp\{-c^0_\phi|t-s|\}$ and DeCAFS is implemented as in (C2)(ii), then for any $\epsilon>0$ (C3) holds with 
\[
\alpha=\max \left\{ 
\frac{c^0_\nu}{c_\nu}\frac{c^0_\phi}{c_\phi}(1+\epsilon), 
\frac{c^0_\nu}{c_\nu}\left(1+\frac{c_\phi}{c^0_\phi}\right)(1+\epsilon), 
\frac{c^0_\eta}{c_\eta}
\right\}
\]
\end{itemize}
\end{proposition}

The following result shows that we can consistently estimate the number of changepoints and gives a bound on the error in the estimate of changepoint locations, if we use DeCAFS under an assumption of a maximum number of changepoints \cite[the assumption of a maximum number changes is for technical convenience, though is common in similar results, e.g.][]{yao1988estimating}.

\begin{theorem} \label{thm:consistency}
Assume data, $y_{1:n}$, is generated as described in (C1), and let $\hat{m}$ and $\hat{\tau}_{1:\hat{m}}$ be the estimated number and location of the changepoints from DeCAFS implemented with parameters given by (C2), penalty $\beta=C\alpha\log n$ for some $C>2$, and a maximum number of changes $m_{\max}\geq m^0$. Then as $n\rightarrow \infty$: if $\phi> 0$ 
\[
\Pr\left(\hat{m}=m^0, \max_{i=1,\ldots,m^0} \left|\hat{\tau}_i-\tau^0_i\right|=0
\right) \rightarrow 1;
\]
and if $\phi=0$
\[
\Pr\left(\hat{m}=m^0, \max_{i=1,\ldots,m^0} \left|\hat{\tau}_i-\tau^0_i\right|\leq (\log n)^2
\right) \rightarrow 1.
\]
\end{theorem}

The most striking part of this result is the very different behaviour between $\phi=0$ and $\phi>0$. In the latter case, asymptotically we detect the position of the changepoints without error. This is because the positive autocorrelation in the noise across the changepoint helps us detect it. In fact, as $n \rightarrow \infty$ the signal for a change at $t$ comes just from the lag-1 difference, $y_{t+1}-y_t$. The variance of $(y_{t+1}-y_t)$ is $O(1/n)$, and its mean is 0 except at changepoints, where it takes a fixed non-zero value. A simple rule based on detecting a change at $t$ if and only if $(y_{t+1}-y_t)^2$ is above some threshold, $c_1(\log n)/n$ for some suitably large constant $c_1$, would consistently detect the changes. For the infill asymptotics we consider, empirically DeCAFS converges to such an approach as $n\rightarrow \infty$.
\section{Simulation Study}
\label{sec:Simulation-Study}

\begin{figure}
    \centering
    \includegraphics[width=0.7\linewidth]{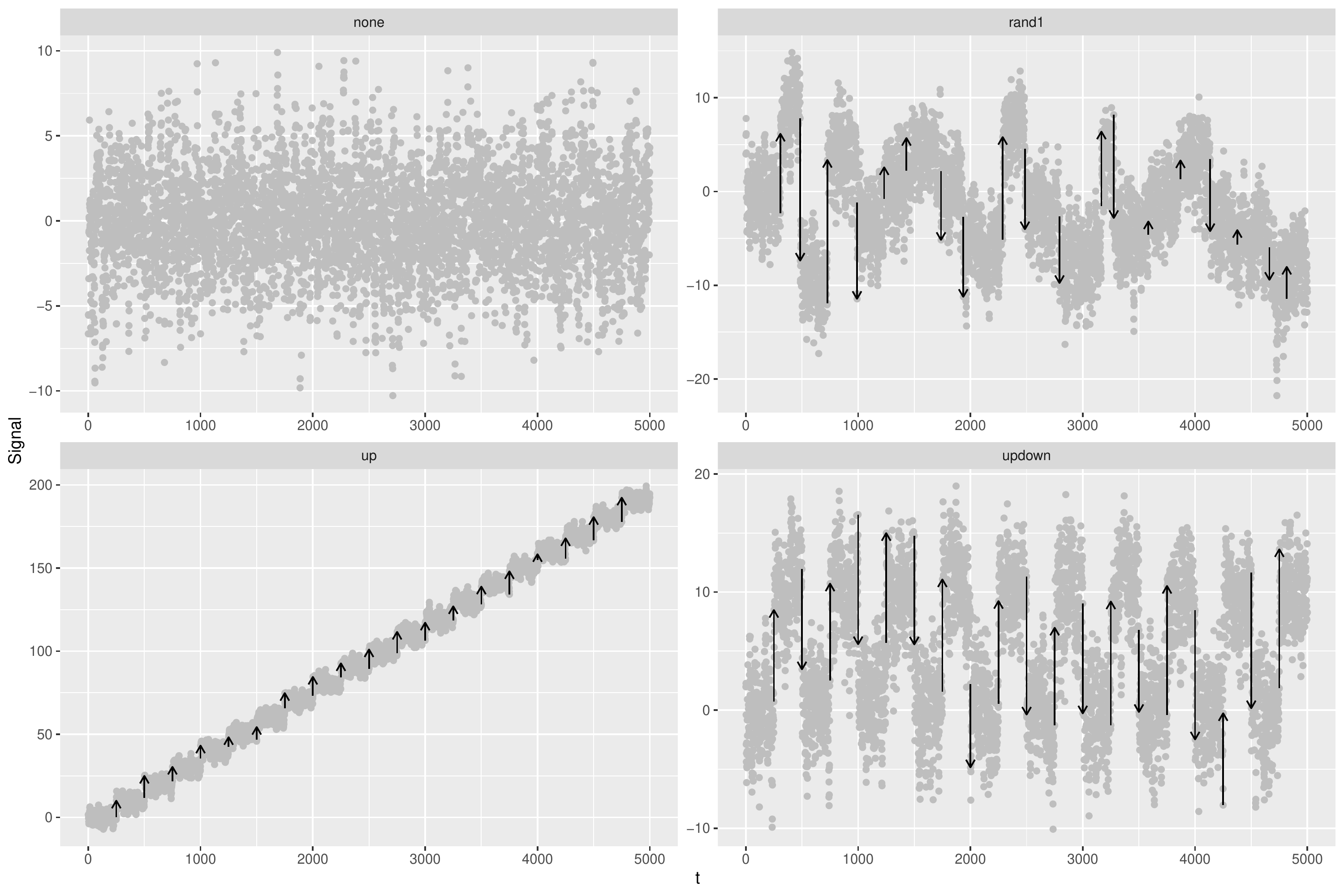}
    \caption{Four different change scenarios. Top-left, no change present, top-right, change pattern with 19 different changes, bottom-left up changes only, bottom-right, up-down changes of the same magnitude. In this particular example data were generated from an AR model with $\phi = 0.7, \ \sigma_\nu = 2$.
    }
    \label{fig:scenarios}
\end{figure}

We now assess the performances of our algorithm in a simulation study on four different change scenarios, illustrated in Figure \ref{fig:scenarios}.

Simulations were performed over a range of evenly-spaced values of $\phi,\ \sigma_\eta,\ \sigma_\nu$.
There are no current algorithms that directly model local fluctuations in the mean, so we compare with two approaches the assume a constant mean between changes: FPOP \cite[]{Maidstone} which also assumes IID noise, and AR1Seg \cite[]{chakar2017robust} that models the noise as an AR(1) process. We compare default implementation of each method, which involves robust estimates of the assumed model parameters. We also compare an implementation of FPOP with an inflated penalty \cite[]{bardwell2019most} to account for the autocorrelated noise. To see the impact of possible misestimation of the model parameters,  we also implement DeCAFS and AR1Seg using the true parameters when this is possible.

We focus on the accuracy of these methods at detecting the changepoints. We deem a predict change as correct if it is within $\pm 2$ observations of a true changepoint. As a measure of accuracy we use the F1 score, which 
is defined as the harmonic mean of the precision (the proportion of detected changes which are correct) and the recall (the proportion of true changes that are detected). The F1 score ranges from 0 to 1, where 1 corresponds to a perfect segmentation. Results reported  are based over 100 replications of each simulation experiment.


\begin{figure}[htp]
	\centering
    \label{fig:Complete_Plot}
    \includegraphics[width = .74\linewidth]{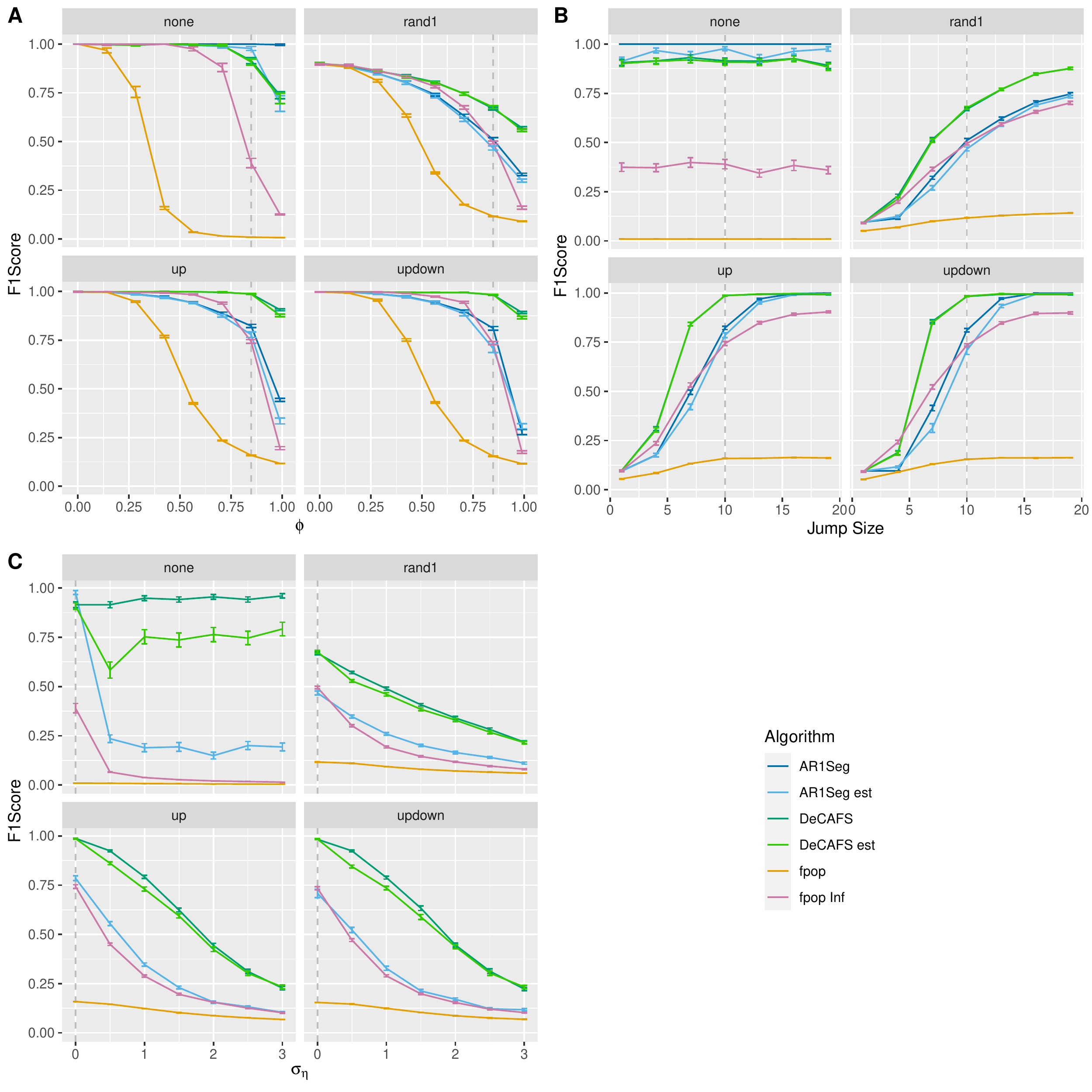}
    \caption{F1 Scores on the 4 different scenarios. In \textbf{A} a pure AR(1) over a range of values of $\phi$, for fixed values of $\sigma_\nu = 2, \ \sigma_\eta = 0$ and a change of magnitude 10. In \textbf{B} a pure AR(1) process with fixed $\phi = 0.85$ and changes in the signal of various magnitudes. In \textbf{C} the full model with $\phi = 0.85$ for a range of values of $\sigma_\eta$. The grey line represent the cross-section between parameters values in \textbf{A}, \textbf{B} and \textbf{C}.
    {AR1Seg est.} and {DeCAFS est.} refer to the segmentation of the relative algorithms with estimated parameters. Note, in \textbf{B} the results from DeCAFS and DeCAFS est overlap so only one line is visible.  
    Other algorithms use the true  parameter values.
    }
\end{figure}

In Figure \ref{fig:Complete_Plot}A we report performances of the various algorithms as we vary $\phi$ for fixed values of $\sigma_{\nu} = 2$ and  $\sigma_\eta = 0$. In Figure \ref{fig:Complete_Plot}B, we additionally fix $\phi = 0.85$, but we vary the size of changes. In these cases there is no random walk component and the model assumed by AR1Seg is correct.

There are a number of conclusions to draw from these results. First we see that the impact of estimating the parameters on the performance of DeCAFS and AR1Seg is small. Second, we see that using a method which ignores autocorrelation but just inflates the penalty for a change does surprisingly well unless the autocorrelation is large, $\phi>0.5$, this is inline with results on the robustness of using a square error cost for detecting changes in mean \cite[]{lavielle2000least}. For high values of $\phi$, DeCAFS is the most accurate algorithm.  The one exception are the simulations where there are no changes: the default penalty choice for AR1Seg is such that it rarely introduces a false positive. 

In Figure \ref{fig:Complete_Plot}C  we explore the effect  of local fluctuations in the mean by varying $\sigma_\eta$.  We see a quick drop off in performance for all methods as $\sigma_\eta$ increases, consistent with the fact that it is harder to detect abrupt changes when the local fluctuations of the mean are greater. Across all experiments, DeCAFS was the most accurate algorithm.

One word of caution when fitting the full DeCAFS model, is that when $\sigma_\eta$ is large it can be difficult to estimate the parameters,  as a model with a very high random walk variance produces data similar to that of a model with constant mean but high autocorrelation. Whilst the impact on detecting changes of any errors when estimating the parameters is small, it can lead to larger errors in the estimate of the signal, $\mu_t$: as different parameter estimates mean that the fluctuations in the data are viewed as either fluctuations in the noise process or in the signal. An example of this is shown in Appendix \ref{App:paramEstim}.

\begin{figure}
    \centering
    \includegraphics[width=.74\linewidth]{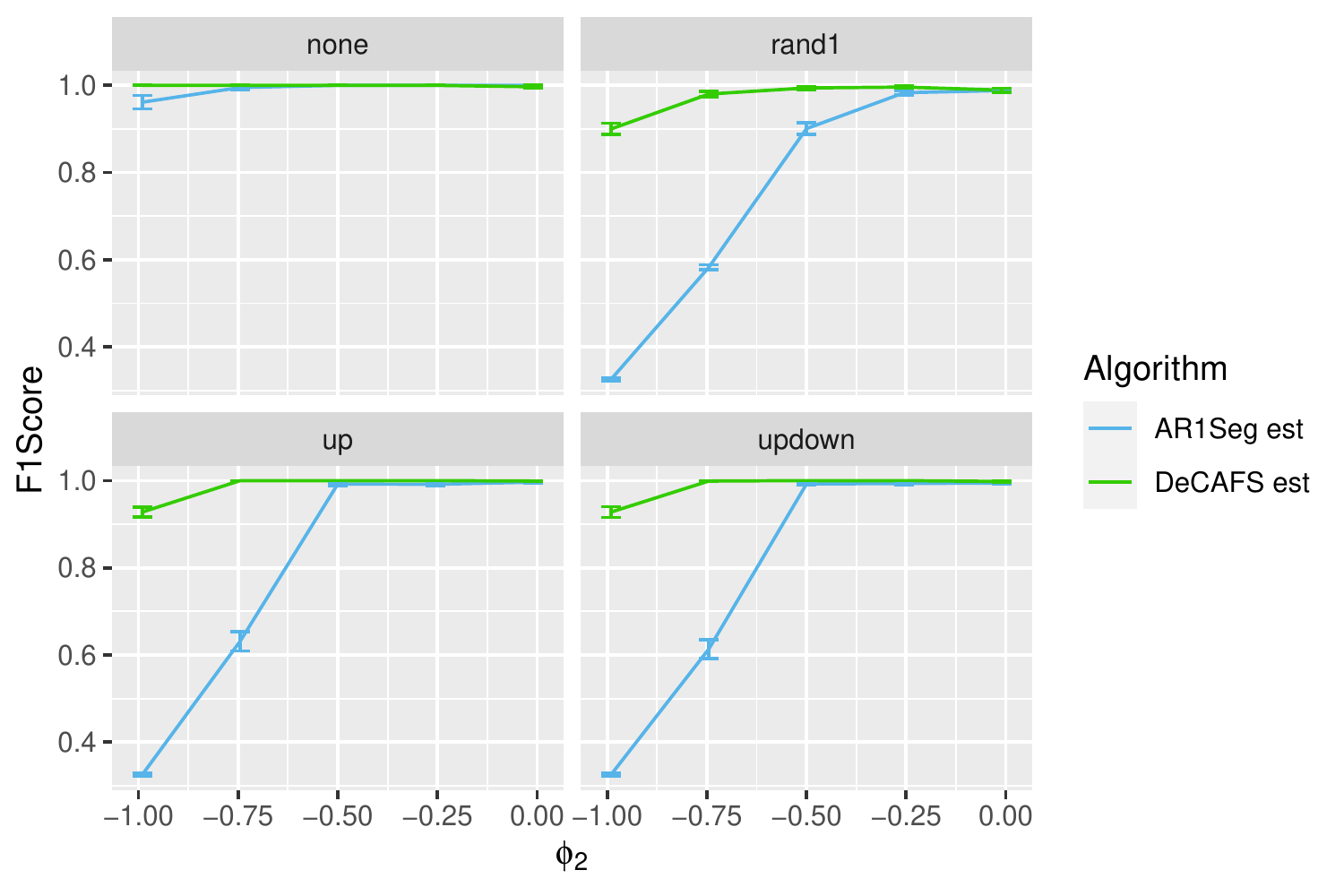}
    \caption{F1 score on different scenarios with AR(2) noise as we vary $\phi_2$. Data simulated fixing $\sigma_{\nu}=2$, $\sigma_\eta=0$ and $\phi_1=0.3$ over a change of size 20.}
    \label{fig:4-AR2}
\end{figure}

Finally we investigate the performance of DeCAFS when its model is incorrect. First we follow \cite{chakar2017robust} and simulate data with a constant mean between changes but with the noise process being AR(2), i.e. $\epsilon_t = \phi_1 \epsilon_{t-1} + \phi_2 \epsilon_{t-2} + \nu_t
$.
In Figure \ref{fig:4-AR2} we report F1 Scores for DeCAFS and AR1Seg as we vary range $\phi_2$. Obviously as $|\phi_2|$ increases, all algorithms perform worse, but the segmentations returned from DeCAFS are the more reliable as we increase the level of model error.

\begin{figure}
    \centering
    \includegraphics[width=.8\linewidth]{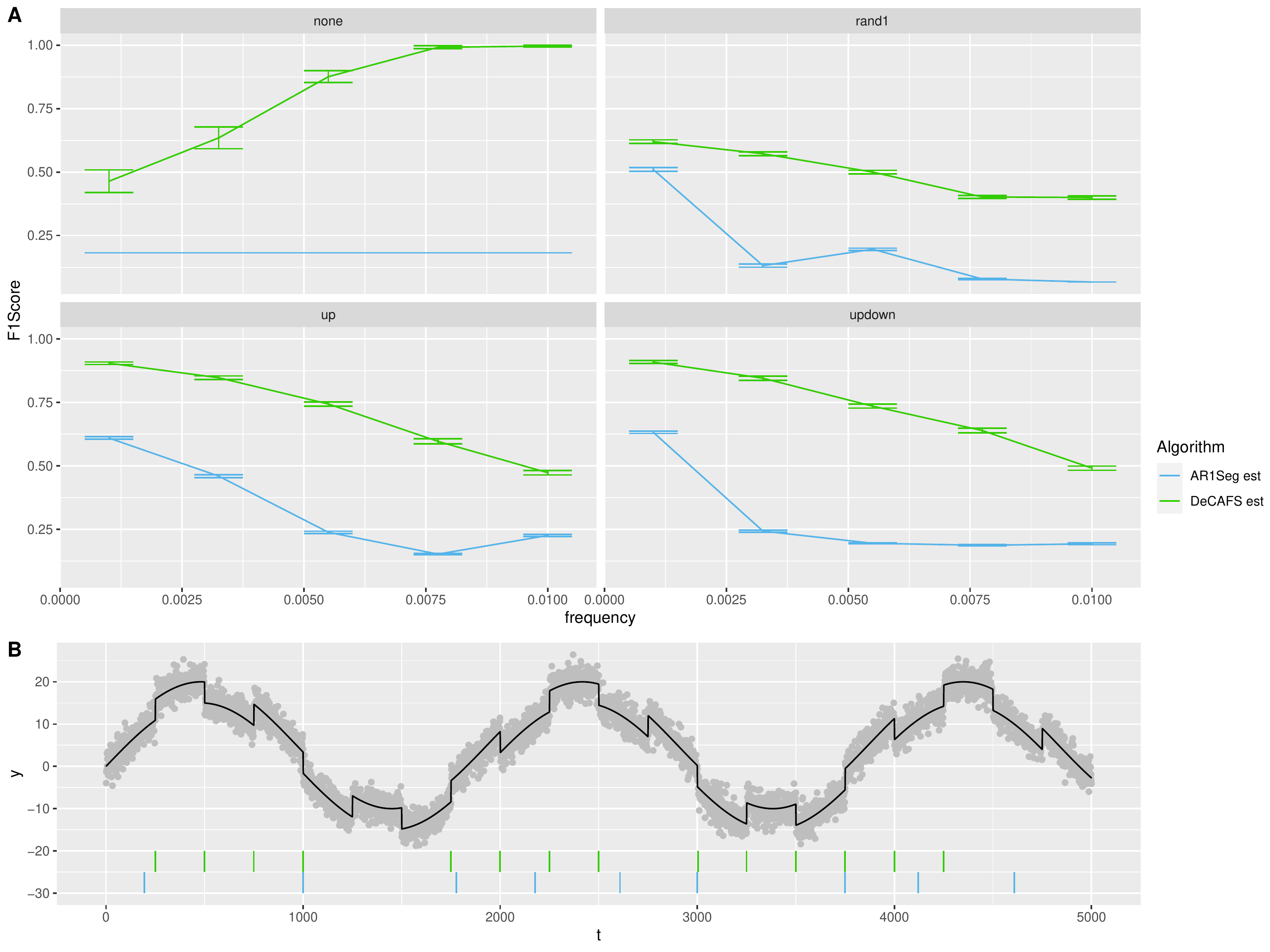}
    \caption{In \textbf{A} the F1Score on the 4 scenarios for the Sinusoidal Model for fixed amplitude of 15, changes of size 5 and IID Gaussian noise with a variance of 4, as we vary the frequency of the sinusoidal process. In \textbf{B} an example of a realization with frequency 0.003 for the updown scenario, vertical segments refer to estimated changepoint locations of DeCAFS (in light green) and AR1Seg (in blue).}
    \label{fig:4-Sinusoidal}
\end{figure}

Second, we consider local fluctuations in the mean that are  generated by a sinusoidal process rather than the random walk model, see Figure \ref{fig:4-Sinusoidal}B.  In Figure \ref{fig:4-Sinusoidal}A we compare performance of DeCAFS and AR1Seg as we vary the frequency of the sinusoidal process. Again we see that DeCAFS gives more reliable segmentations in these cases. In the three change scenarios  performance decrease as we increase the frequency of the process. In these cases it becomes significantly harder to detect any changepoints, however DeCAFS still has higher scores than AR1Seg since it is more robust and returns fewer false positives. 

For the no change scenario, interestingly, we observe an increase in DeCAFS performances: for low frequencies, in roughly half of the simulations, the estimated parameters used by DeCAFS  correspond to incorrectly modelling the process as a pure AR(1) process (\textit{i.e}. $\hat{\phi} \ne 0, \ \hat{\sigma}_\eta = 0$) which results in an increased number of false positives. If we knew that the noise was independent we could overcome this problem by enforcing $\phi = 0$.

\section{Gene Expression in \textit{Bacilus subtilis}}\label{sec:Real-Data}

We now evaluate DeCAFS on estimating the expression of cells in the bacteria \textit{Bacilus subtilis}. Specifically we analyze data from  \cite{nicolas2009transcriptional}, which is  data from tiling arrays with a resolution of less than 25 base pairs. The array contains several hundred thousand probes which are ordered according to their position on the bacterial chromosome.
For a probe, labelled $t$ say, we get an RNA expression measure, $Y_t$. Figure \ref{fig:data_bacillus} shows data from 2000 probes.  
\ifnotblind
Code and data used in our analyses, presented below, are available on forgemia : \url{https://forgemia.inra.fr/guillem.rigaill/l2fpop_tiling_array_data}.
\else 
Code and data used in our analyses, presented below, are available at : \url{https://***}.
\fi

\begin{figure}
    \centering
    \includegraphics[width=16cm]{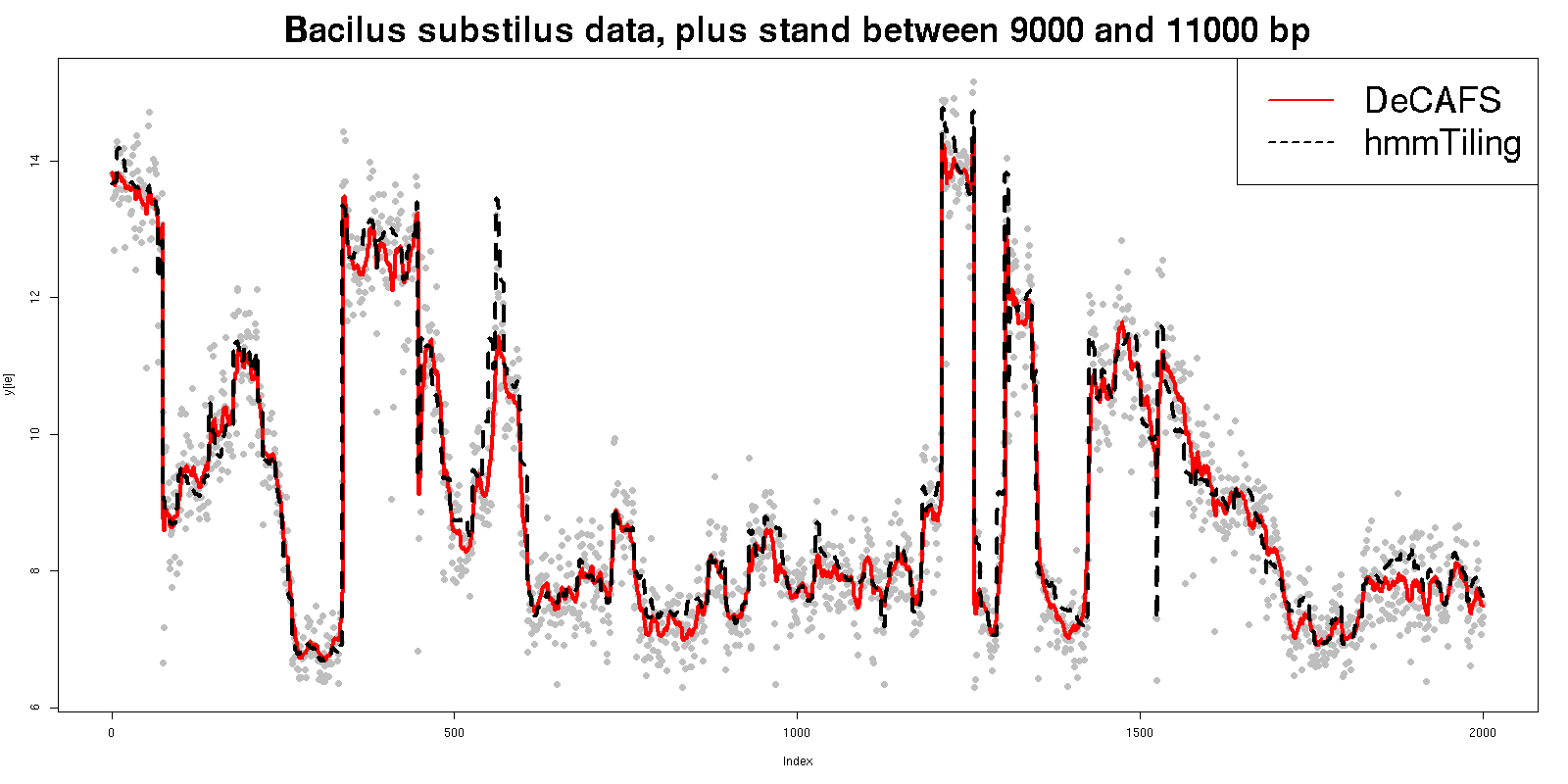}
    \caption{Data on 2000 bp of the plus-strand of the \textit{Bacilus subtilis} chromosome. Grey dots show the original data. The plain red line represents the estimated signal of DeCAFS with a penalty of $10 \log(n)$. The dashed black line represents the estimated signal of hmmTiling.}
    \label{fig:data_bacillus}
\end{figure}

The underlying expression level is believed to undergo two types of transitions, large changes which \cite{nicolas2009transcriptional} call shifts and small changes which they call drifts. Thus it naturally fits our modelling framework of abrupt changes, the shifts, between which there are local fluctuations caused by the drifts. To evaluate the performance of DeCAFS at estimating how the gene expression levels vary across the genome we will compare to the hmmTiling method of \cite{nicolas2009transcriptional}. This method fits a discrete state hidden Markov model to the data, with the states being the gene expression level, and the dynamics of the hidden Markov model corresponding to either drifts or shifts. As a comparison of computational cost for of the two methods, DeCAFS takes about 7 minutes to analyse data from one of the strands, each of which contains around 192,000 data points. \cite{nicolas2009transcriptional} reported a runtime of 5 hours and 36 minutes to analyse both strands.

 A comparison of the estimated gene expression level from DeCAFS and from hmmTiling, for a 2000 base pair region of the genome, is shown in Figure \ref{fig:data_bacillus}. We see a close agreement in the estimated level for most of the region, except for a couple of regions where hmmTiling estimates abrupt changes in gene expression level that DeCAFS does not.

To evaluate which of DeCAFS and hmmTiling is more accurate, we follow \cite{nicolas2009transcriptional} and  see how well the estimated gene expression levels align with bioinformatically predicted promoters and terminators.
A promoter roughly corresponds to the start of a gene, and a terminator the end, and we expect gene expression to increase around a promoter and decrease around a terminator. 

For promoters, consider all probe locations $t$ from the tiling chip and consider a threshold parameter $\delta$.
We can count the number of probe locations with a predicted difference $ \hat{d}_t = \hat{\mu}_{t + 1} - \hat{\mu}_t $ strictly greater than $\delta$. We call this 
$R(\delta).$ Among those probes, we can count how many have a promoter nearby (within 22 base pairs). We call this $M(\delta)$. By symmetry we can define an equivalent measure for terminators. A method is better than another if for the same $R(\delta)$ it achieves a larger $M(\delta).$ 

\begin{figure}
    \centering
    \includegraphics[width=8cm]{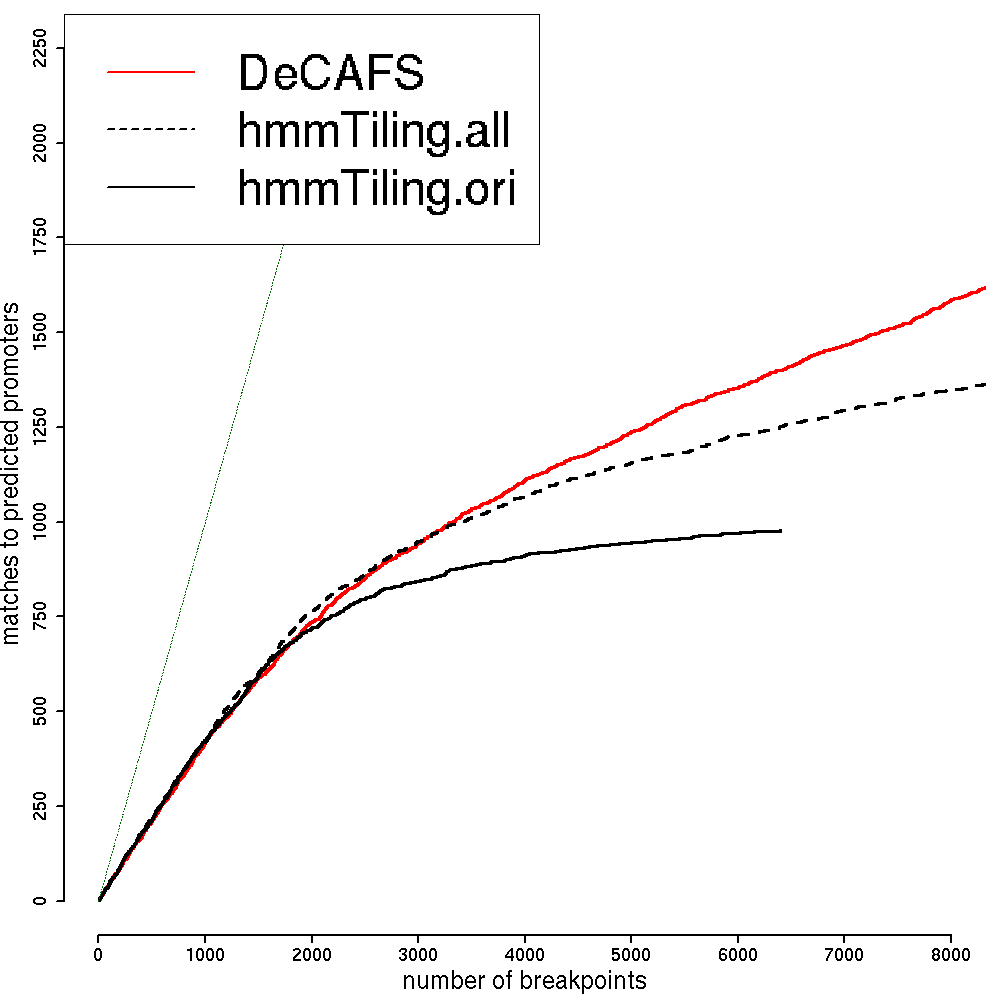}
    \includegraphics[width=8cm]{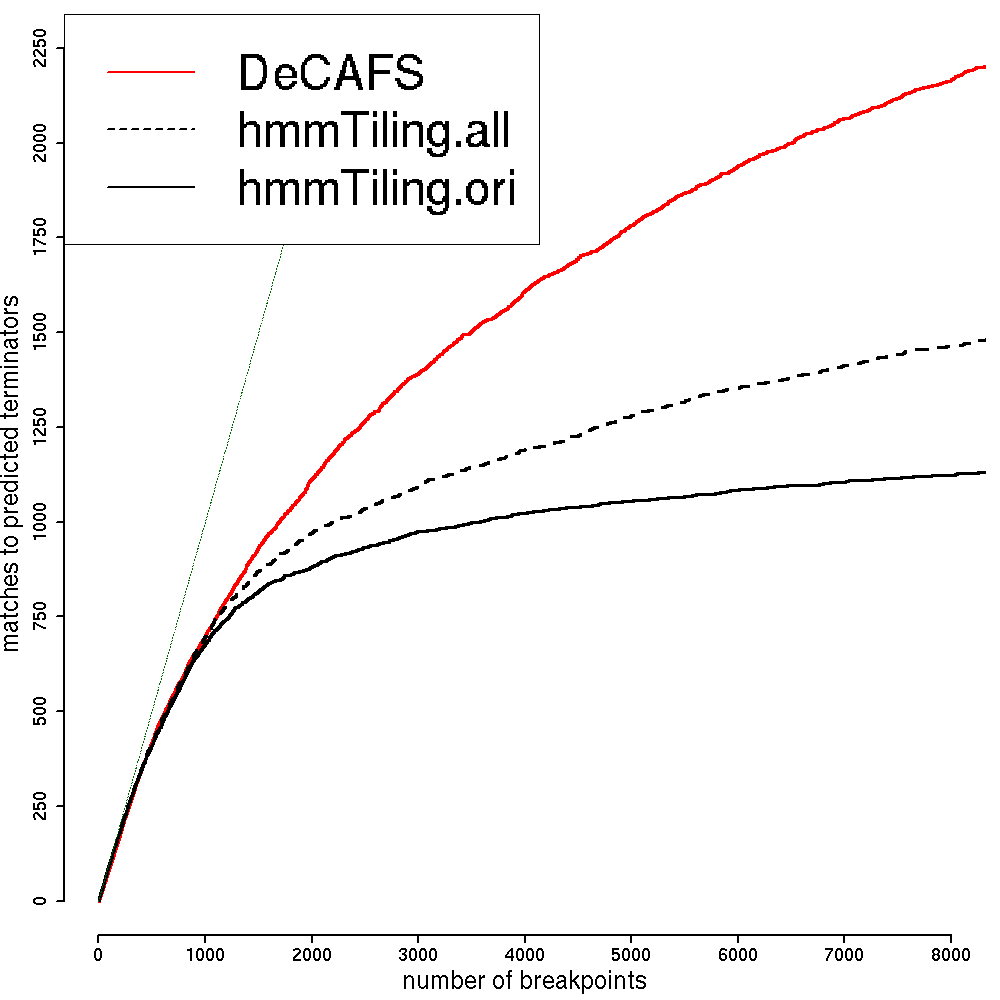}
    \caption{Benchmark comparisons. The number of promoters (left) and terminators (right) correctly predicted, $M(\delta)$ using a 22 bp distance cutoff, as a function of the number of predicted breakpoints, $R(\delta)$. Plain black lines are the results of hmmTiling \cite[as reported in Figure 4 of][]{nicolas2009transcriptional}). 
    Dotted black lines are the results of hmmTiling when considering all probes rather than only those called transitions. Plain red lines are the results of DeCAFS using $\beta=10 \log(n)$. 
    The thin dark-green leaning line represent $y=x.$}
    \label{fig:hmmtiling}
\end{figure}

Figure \ref{fig:hmmtiling} plots $M(\delta)$ against $R(\delta)$ as we vary $\delta$ for DeCAFS and two different estimates from hmmTiling. 
In the case of promoters the prediction of hmmTiling is slightly better than DeCAFS for  lower thresholds but noticeably worse for higher thresholds.  In the case of terminators the prediction of DeCAFS are clearly better than those of hmmTiling. Given that DeCAFS was not developed to analyze such data we believe that its relatively good performances for promoters and better performances for terminators is a sign of its versatility.

\section{Discussion}

There are various ways of developing the DeCAFS algorithm, that build on other extensions of the functional pruning version of optimal partitioning. For example, to make the method robust to outliers,  we can use robust losses, such as the bi-weight loss, instead of square error loss to measure our fit to the data \cite[]{Fearnhead}. Alternatively we can incorporate additional constraints on the underlying mean such as monotonicity \cite[]{hocking2017log} or that the mean decays geometrically between changes \cite[]{jewell2018exact,jewell2018fast}. Finally, the algorithm is inherently sequential and thus should be straightforward to adapt to an online analysis of a data stream.

We do not claim that the method we present in Section \ref{sec:parameter-estimation} for estimating the parameters in our model is best. It is likely that more efficient or more robust methods are possible, for example using different robust estimates of the variances of the $k$-lag difference data \cite[]{rousseeuw1993alternatives}; or using iterative procedures where we estimate the changepoints, and then conditional on these changepoints re-estimate the parameters. Using better estimates should lead to further improvement on the statistical performance we observed in Section \ref{sec:Simulation-Study}. Our theoretical results suggest that for estimating changes, mis-estimation of the parameters, or errors in our model for the noise or local fluctuations, can be corrected by inflating the penalty for adding a changepoint. As such, in applications we would suggest implementing the method for a range of penalty values, for example using the CROPS algorithm \cite[]{Haynes}, and then choosing the number of penalties using criteria that consider how the fit to data improves as we add more changes \cite[e.g.][]{arlot_etal_capushe2016,fryzlewicz2018detecting,arlot2019minimal}.

\noindent
{\bf Acknowledgements}
\ifnotblind
This work was supported by EPSRC grant EP/N031938/1, and an ATIGE grant from Genopole. The IPS2 benefits from the support of the LabEx Saclay Plant Sciences-SPS.
\fi
We thank Pierre Nicolas for providing the output of hmmTiling on the \textit{Bacillus subtilis} data and his R code allowing us to generate Figure \ref{fig:hmmtiling}, which closely resembles Figure 4 of \cite{nicolas2009transcriptional}. 


\bibliographystyle{agsm}
\bibliography{Bibliography}
\newpage
\ifJASA
\setcounter{page}{1}
\fi

\appendix 

\begin{center}
{\large\bf SUPPLEMENTARY MATERIAL}
\end{center}

\section{Proof of Proposition \ref{th:propositionNEW}} 
\label{App:DP_Proof}

The initial condition for $Q_1(\mu)$ follows immediately from its definition.

Then, for $t \in \{2, ..., n\}$, we need to condition the problem separately on whether or not we have a changepoint. If we consider no change in the mean of the signal, then we can we can re-arrange the cost at time $t$ based on the cost at time $t - 1$ in the following way:
\begin{align*}
	Q_t(\mu| \delta_t = 0)
	&= \min_{\substack{u }} \left\{ Q_{t-1}(u) + \lambda (\mu - u) ^ 2 + \gamma \Big((y_t - \mu) - \phi (y_{t-1} - u)\Big)^2 \right\}\,.\\
\end{align*}
Similarly, when we have a change:
\begin{align*}
Q_t(\mu | \delta_t \neq 0) &= \min_{\substack{u, \delta}} \left\{ Q_{t-1}(u) + \lambda (\mu - u - \delta) ^ 2 + \gamma \Big((y_t - \mu) - \phi (y_{t-1} - u)\Big)^2 + \beta \right\}\\
&= \min_{\substack{u }} \left\{ Q_{t-1}(u) +  \gamma \Big((y_t - \mu) - \phi (y_{t-1} - u)\Big)^2 + \beta \right\}\\
\end{align*}
where the second equality comes from minimising over $\delta$.

Lastly, to obtain the whole cost at time $t$ we take the minimum of these two functions:
\begin{align*}
	Q_t(\mu) &= \min \left\{Q_t(\mu | \delta_t = 0),  Q_t(\mu | \delta_t \neq 0) \right\} \\
	& = \min_{\substack{u}} \left\{ Q_{t-1}(u) + \min\{\lambda (\mu - u) ^ 2, \beta\} + \gamma \Big((y_t - \mu) - \phi (y_{t-1} - u)\Big)^2 \right\}\,.
\end{align*}

\hfill $\Box$

\section{Proof of Proposition \ref{th:propositionALGO}} 
\label{App:Cost_Computation}

From the result obtained in Appendix \ref{App:DP_Proof}, simple, albeit tedious, algebraic manipulation enables us to re-write the recursions for $Q_t(\mu | \delta_t \neq 0)$  and $Q_t(\mu | \delta_t = 0)$ in terms of the infimal convolution operator. Let $z_t=y_t-\phi y_{t-1}$.

For $Q_t(\mu | \delta_t \neq 0)$, we can rearrange
\begin{eqnarray*}
 \lefteqn{\gamma \Big( (y_t-\mu)-\phi(y_{t-1}-u) \Big)^2 
=  \gamma (z_t-\mu+\phi u )^2} \\
&=& \gamma (z_t-\mu)^2 +\gamma\phi^2 u^2 + 2\gamma\phi u z_t -2\gamma\phi u \mu \\
&=& \gamma (z_t-\mu)^2 +\gamma\phi^2 u^2 + 2\gamma\phi u z_t + \gamma\phi(u-\mu)^2 -\gamma\phi u^2 - \gamma\phi \mu^2  \\
&=&\gamma\phi(u-\mu)^2- \gamma\phi(1-\phi)\left(u- \frac{z_t}{1-\phi} \right)^2 +\gamma\phi \frac{z_t^2}{1-\phi} + \gamma (z_t-\mu)^2  - \gamma\phi \mu^2 
\end{eqnarray*}
Hence, we have

\begin{eqnarray*}
Q_t(\mu | \delta_t \neq 0)&=&\min_{u \in \mathbb{R}} \left[ Q_{t-1}(u) - \gamma\phi(1-\phi)\left(u- \frac{z_t}{1-\phi} \right)^2
+\gamma\phi (u-\mu)^2 \right]\\
& &+ \frac{\gamma}{1-\phi} (z_t - (1-\phi)\mu)^2 +\beta \\
&=& \mathsf{INF}_{\mathbb{Q}_{t-1},\gamma\phi}(\mu) +  \frac{\gamma}{1-\phi} \Big( z_t - (1-\phi)\mu\Big)^2 +\beta  = Q^{\neq}_{t}(\mu),
\end{eqnarray*}

where 
\[
\mathbb{Q}_{t-1}(u)=Q_{t-1}(u)- \gamma\phi(1-\phi)\left(u- \frac{z_t}{1-\phi} \right)^2.
\]

Similar, for $Q_t(\mu | \delta_t = 0)$, we can rearrange
\begin{eqnarray*}
\lefteqn{\lambda (\mu-u)^2 + \gamma \Big( (y_t-\mu)-\phi(y_{t-1}-u) \Big)^2}  & & \\
&=& (\gamma\phi+\lambda)(u-\mu)^2- \gamma\phi(1-\phi)\left(u- \frac{z_t}{1-\phi} \right)^2 +\gamma\phi \frac{z_t^2}{1-\phi} + \gamma (z_t-\mu)^2  - \gamma\phi \mu^2.
\end{eqnarray*}
Hence
\begin{eqnarray*}
Q_t(\mu | \delta_t = 0)
&=& \mathsf{INF}_{\mathbb{Q}_{t-1},\gamma\phi+\lambda}(\mu) +  \frac{\gamma}{1-\phi} \Big( z_t - (1-\phi)\mu\Big)^2  = Q^{=}_{t}(\mu),
\end{eqnarray*}
where $\mathbb{Q}_{t-1}$ is defined above.

\hfill $\Box$

\section{Proof of Theorem  \ref{theorem}} \label{App:InfProof}

The proof is based on the following lemmas. 
\begin{lemma}
\label{L1}
For any lower-bounded function $Q : \R \to \R$, we define the proxy operator
$$\hat{u}_{\omega} : \left\{
    \begin{array}{l}
        \R \to \R\\
        \theta \,\,\mapsto\,\, \min\Big\{\underset{u\in \R}{\argmin} \Big(Q(u) + \omega(u-\theta)^2 \Big)\Big\}\,.
    \end{array}
\right.$$
The function $\hat{u}_{\omega}$ is non-decreasing on $\R$.
\end{lemma}

Notice that we use a minimum in the definition of $\hat{u}_{\omega}$ only to get a single-valued function (we could have done another choice). Indeed, taking $Q = \min(q_1,q_2)$ with $q_1(\theta) = (\theta+1)^2$ and $q_2(\theta) = (\theta-1)^2$, we have $\hat{u}_1(0) = \underset{u\in \R}{\argmin}(Q(u) + u^2) = \{-\frac{1}{2}, \frac{1}{2}\}$ and we need to make a choice (here the smallest value) to get a well-defined function.

{\bf Proof:}
We consider $\theta_1,\theta_2 \in \R$ such that $\theta_1 < \theta_2$ and define $\hat{u}_1 = \hat{u}_{\omega}(\theta_1)$, $\hat{u}_2= \hat{u}_{\omega}(\theta_2)$. Using the definition of $\hat{u}_1$ and $\hat{u}_2$ we can write
$$Q(\hat{u}_1) + \omega(\hat{u}_1-\theta_1)^2 \le Q(\hat{u}_2) + \omega(\hat{u}_2-\theta_1)^2\,,$$
$$Q(\hat{u}_2) + \omega(\hat{u}_2-\theta_2)^2 \le Q(\hat{u}_1) + \omega(\hat{u}_1-\theta_2)^2\,.$$
Summing the two inequalities, the $Q$ terms cancel out and we get
$$(\hat{u}_2-\hat{u}_1)(\theta_2-\theta_1) \ge 0\,,$$
which shows that $\hat{u}_1 \le \hat{u}_2$ and the result is proven.
\hfill $\Box$

In our stochastic models
the function $Q$ is described by a list of functions $Q = (q_1,...,q_s)$ with $Q\restrict{D_i} = q_i$  where $D_i = [d_i,d_{i+1}[ \subset \R$ is an interval and $\{D_i\}_{ i = 1,...,s}$ a partition of the real line. To compute the convolution, we define the functions 
$$
\overline{q}_i(u) = \left\{
    \begin{array}{ll}
        q_i(u) \,,\quad &\hbox{if}\quad u \in D_i\,,\\
         +\infty \,,\quad &\hbox{if}\quad u \not\in D_i\,.\\
    \end{array}
\right.
$$
The infimal convolution of this kind of functions can be analytically described.

\begin{lemma}
\label{L2}
The infimal convolution of a function $\overline{q}$ given by
$$
\overline{q}(u) = \left\{
    \begin{array}{ll}
        q(u) \,,\quad &\hbox{if}\quad u \in [m_1,m_2]\,,\\
         +\infty \,,\quad &\hbox{if}\quad u \not\in [m_1,m_2]\,,\\
    \end{array}
\right.
$$
with any function $q$ continuously differentiable ($C^1$) on $[m_1,m_2]$ is given by
\begin{equation}
\label{Fstar}
\mathsf{INF}_{\overline{q},\omega}(\theta) = \left\{
    \begin{array}{ll}
    \underset{u\in [m_1,m_2]}{\min} \Big(q(u) + \omega(u-\theta)^2\Big)\,,\quad &\hbox{if}\quad \theta \in [m_1^*,m_2^*]\,,\\
        q(m_1) + \omega(m_1-\theta)^2\,, \quad &\hbox{if}\quad \theta < m_1^*\,,\\
         q(m_2) + \omega(m_2-\theta)^2\,, \quad &\hbox{if}\quad \theta > m_2^*\,,\\
    \end{array}
\right.
\end{equation}
with $[m_1^*,m_2^*] = [\frac{1}{2\omega}q'(m_1) + m_1, \frac{1}{2\omega}q'(m_2) + m_2]$.

\end{lemma}
{\bf Proof:}
Using Lemma \ref{L1} we know that the proxy operator $\hat{u}_{\omega}$ with $Q = \overline{q}$ is a non-decreasing function in $\theta$. Thus, there exist $m^*_1, m^*_2 \in \R$ such that for all $\theta \in [m^*_1,m^*_2]$, the argminimum of $\overline{q}_{\omega} : u \mapsto \overline{q}(u) + \omega(u-\theta)^2$ belongs to the interval $[m_1,m_2]$ and $\overline{q} = q$ on this interval. As $q$ is $C^1$, the stationary points of $\overline{q}_{\omega}$ are solutions of the equation $\frac{1}{2\omega}q'(u) + u = \theta$. At point $m_1$ (resp. $m_2$) we have the argminimum $m_1^*$ with $m_1^* = \frac{1}{2\omega}q'(m_1) + m_1$ (resp. $m_2^* = \frac{1}{2\omega}q'(m_2) + m_2$). If we have $\theta < m^*_1$, then the argminimum of $\overline{q}_{\omega}$ is less than $m_1$ and then attained at $u = m_1$ (as $\overline{q}(u) = +\infty$ if $u < m_1$) and we get $\mathsf{INF}_{\overline{q},\omega}(\theta) = q(m_1)+\omega(m_1-\theta)^2$. With the same reasoning in case $ \theta > m^*_2$ the lemma is proven.
\hfill $\Box$

Using these two lemmas, we can prove the following proposition.
\begin{proposition}
    \label{prop}
    The infimal convolution of the functional cost $Q = (q_1,...,q_s)$ is given by $\mathsf{INF}_{Q,\omega} = (\mathsf{INF}_{q_1,\omega},...,\mathsf{INF}_{q_s,\omega})$.
\end{proposition}

{\bf Proof:}
    With previously introduced notations we have $Q(\theta) = \underset{i=1,...,s}{\min}\{\overline{q}_i(\theta)\}$. Then 
    $$\mathsf{INF}_{Q,\omega}(\theta) = \min_{u\in \mathbb{R}}\left(\underset{i=1,...,s}{\min}\{\overline{q}_i(\theta)\} + \omega(u-\theta)^2 \right) = \min_{u\in \mathbb{R}}\left(\underset{i=1,...,s}{\min}\{\overline{q}_i(\theta) + \omega(u-\theta)^2 \}\right)$$
    $$= \underset{i=1,...,s}{\min}\left\{\min_{u\in \mathbb{R}}\left(\overline{q}_i(\theta) + \omega(u-\theta)^2\right) \right\}\,,$$
    which gives us $\mathsf{INF}_{Q,\omega}(\theta) = \underset{i=1,...,s}{\min}\{\mathsf{INF}_{\overline{q}_i,\omega}(\theta)\}$ for all $\theta \in \R$. $\mathsf{INF}_{Q,\omega}$ can be described by a list $(\mathsf{INF}_{\overline{q}_{\nu(1)},\omega},\mathsf{INF}_{\overline{q}_{\nu(2)},\omega},...,\mathsf{INF}_{\overline{q}_{\nu(r)},\omega})$ with $\nu(i) \in \{1,...,s\}$. The function $i \mapsto \nu(i)$ is increasing due to Lemma \ref{L1} (and $\nu(r) = s$).
\hfill $\Box$

In order to prove Theorem \ref{theorem} we only need to show  that we can remove the overline sign in $(\mathsf{INF}_{\overline{q}_{\nu(1)},\omega},\mathsf{INF}_{\overline{q}_{\nu(2)},\omega},...,\mathsf{INF}_{\overline{q}_{\nu(r)},\omega})$ without consequences.  We assume that $Q$ is continuously differentiable ($C^1$) except at the points $d_i$ for $i=2,...,s$. The left and right derivatives at point $\theta$ are respectively designated by $Q'_{-}(\theta)$ and $Q'_{+}(\theta)$. With these assumptions we can prove the following result.

\begin{lemma}
    \label{L3}
    If at points $\theta = d_i$ we have $Q'_{-}(d_i) > Q'_{+}(d_i)$ then $d_i$ is never an argminimum for the convolution.
\end{lemma}

{\bf Proof:}
    We study the stationary points of $Q_{\omega} : u \mapsto Q(u) + \omega(u-\theta)^2$. The necessary condition for optimality $Q_{\omega}(u) \le Q_{\omega}(u+\epsilon)$ for all $\epsilon$ leads to the inequalities
    $$\frac{1}{2\omega}Q'_{-}(u) + u \le \theta \le \frac{1}{2\omega}Q'_{+}(u) + u\,.$$
    In case $Q'_{-}(u)>Q'_{+}(u)$ there exists no such $\theta$ satisfying the two inequalities so that this $u$ can not be used in any minimization of $Q_{\omega}$ and $\hat{u}_{\omega}$ never takes this value.
\hfill $\Box$

With this result the $d_i$ never appear as an argminimum for the convolution and using Lemma \ref{L2}, we get $(\mathsf{INF}_{\overline{q}_{\nu(1)},\omega},\mathsf{INF}_{\overline{q}_{\nu(2)},\omega},...,\mathsf{INF}_{\overline{q}_{\nu(r)},\omega}) = (\mathsf{INF}_{q_{\nu(1)},\omega},\mathsf{INF}_{q_{\nu(2)},\omega},...,\mathsf{INF}_{q_{\nu(r)},\omega})$ in Proposition \ref{prop}. \\

By looking at updates in Propositions \ref{th:propositionNEW} and \ref{th:propositionALGO}, it remains to prove that at any time step, no slope discontinuity at $\theta = d$ in $Q_t=Q$ satisfies the inequality $Q^{'}_{-}(d)<Q^{'}_{+}(d)$. We prove this result by recursion: at the initialisation step, there is no such breakpoint in the cost function and all the min operators involved can not produce them. We eventually have to prove that the infimal transformation in Lemma \ref{L2} can not introduce these discontinuities.

Around $m_1^*$ in (\ref{Fstar}) we have:

$$
\frac{d}{d\theta} \mathsf{INF}_{Q,\omega}(\theta)  = \left\{
    \begin{array}{ll} \frac{d \hat{u}_\omega(\theta)}{d\theta} q'(\hat{u}_\omega(\theta)) + 2\omega(\frac{d \hat{u}_\omega(\theta)}{d\theta} - 1)(\hat{u}_\omega(\theta)-\theta)\,,\quad &\hbox{if}\quad \theta  \ge m_1^*\,,\\
        -2\omega(m_1-\theta)\,, \quad &\hbox{if}\quad \theta < m_1^*\,,\\
    \end{array}
\right.$$
with the function $\theta \mapsto \hat{u}_{\omega}(\theta)$ being the argminimum of the infimal convolution (see Lemma \ref{L1}). By direct computation with $\hat{u}_{\omega}(m_1^*) = m_1$ and $m^*_1 = \frac{1}{2\omega}q'(m_1) + m_1$ we get $\mathsf{INF}^{'}_{Q,\omega -}(m_1^*) = q'(m_1) = \mathsf{INF}^{'}_{Q, \omega +}(m_1^*)$. This result achieves the proof of Theorem \ref{theorem}.

\section{Proofs for Section \ref{sec:Theory} }\label{App:Theory}

By definition of the random-walk model for $\tilde{\eta}_{1:n}$ in Equation (\ref{eq:RWmodel}) and the auto-regressive model for $\epsilon_{1:n}$ in Equation (\ref{eq:ARmodel}) we have that the covariance matrices have entries
\[
[\SAR]_{ij}=\frac{\sigma^2_\nu}{1-\phi^2} \phi^{|i-j|}, ~~~~ [\SRW]_{ij}= \sigma^2_{\eta} \min\{i,j\}.
\]
It is straightforward to find that their inverses have entries
\[
[\SAR^{-1}]_{ij}=\left\{\begin{array}{cl} 1/\sigma^2_\nu & \mbox{if }i=j=1\mbox{ or } n,\\
(1+\phi^2)/\sigma^2_\nu & \mbox{if }i=j\neq 1\mbox{ or } n,\\
-\phi/\sigma^2_\nu & \mbox{if } |i-j|=1,\\
0& \mbox{otherwise}, \end{array} \right.
\]
and
\[
[\SRW^{-1}]_{ij}=\left\{\begin{array}{cl} 1/\sigma_\eta^2 & \mbox{if }i=j= n,\\
2/\sigma^2_\eta & \mbox{if }i=j \neq n,\\
-1/\sigma^2_\eta & \mbox{if } |i-j|=1,\\
0& \mbox{otherwise}, \end{array} \right.
\]

The unpenalised cost conditional on the set of changepoints is
\begin{eqnarray*}
 \lefteqn{\mathcal{C}(\tau_{1:m})  = \min\left\{ 
       (1-\phi^2)\gamma(y_1-\mu_1)^2+ \sum_{t = 2}^n  \left[ \lambda(\mu_{t}  -\mu_{t-1} - \delta_{t}) ^ 2 
         + \gamma \Big((y_t - \mu_t) - \phi (y_{t-1} - \mu_{t-1})\Big)^2 \right] \right\} } \\
&=& \min 
\left\{ 
       (1-\phi^2)\gamma(y_1-\mu_1)^2 +\sum_{t = 2}^n  \left[ \lambda (\tilde{\eta}_t-\tilde{\eta}_{t-1}) ^ 2 
         + \gamma \Big((y_t - \mu_t) - \phi (y_{t-1} - \mu_{t-1} )\Big)^2 \right] \right\}
\end{eqnarray*}
where the minimisation is over $\mu_{1:n}$, and $\delta_{2:n}$ consistent with the set of changepoints; and we have made a change of variables such that $\tilde{\eta}_i-\tilde{\eta}_{i-1}=\mu_i-\mu_{i-1}-\delta_i$ for $i=2,\ldots,n$ in the second equality. 

This change of variables is not unique, and we get the same value for any choice of $\tilde{\eta}_1$. Thus we trivially have that
\begin{eqnarray*}
\lefteqn{\mathcal{C}(\tau_{1:m})} \\
&=& \min 
\left\{ 
       (1-\phi^2)\gamma(y_1-\mu_1)^2 +\sum_{t = 2}^n  \left[ \lambda (\tilde{\eta}_t-\tilde{\eta}_{t-1}) ^ 2 
         + \gamma \Big((y_t - \mu_t) - \phi (y_{t-1} - \mu_{t-1} )\Big)^2 \right] +\lambda \tilde{\eta}_1^2\right\},
\end{eqnarray*}
where the minimisation is now also over $\tilde{\eta}_1$, and the minimum is attained with $\tilde{\eta}_1=0$.

By our definition of the matrix $X_{\tau_{1:m}}$ we have that if $\Delta=(\mu_1-\tilde{\eta}_1,\delta_{\tau_{1:m}})$ we can write $\mu_{1:n}=X_{\tau_{0:m}}\Delta+\tilde{\eta}_{1:n}$. Thus by re-writing the sums, e.g. 
\[
\sum_{t = 2}^n \lambda \{\tilde{\eta}_t-\tilde{\eta}_{t-1}\} ^ 2 +\lambda \tilde{\eta_1}^2 = \tilde{\eta}_{1:n}^T \SRW^{-1} \tilde{\eta}_{1:n},
\]
as $\lambda=1/\sigma^2_\eta$, gives that 
\begin{equation} \label{eq:cost1}
\mathcal{C}(\tau_{1:m})= \min_{\Delta,\tilde{\eta}_{1:n}} \left[ (y_{1:n}-X_{\tau_{0:m}}\Delta-\tilde{\eta}_{1:n})^T \SAR^{-1} (y_{1:n}-X_{\tau_{0:m}}\Delta-\tilde{\eta}_{1:n})+\tilde{\eta}_{1:n}^T\SRW^{-1}\tilde{\eta}_{1:n}
\right].
\end{equation}

\noindent {\bf Proof of Proposition \ref{prop:null}.}
To simplify notation we will write $\tilde{\eta}$ for $\tilde{\eta}_{1:n}$, $y$ for $y_{1:n}$ and $X$ for $X_{\tau_{0:m}}$. 
Re-writing right-hand side of (\ref{eq:cost1}) gives
\begin{eqnarray*}
\lefteqn{\min_{\Delta,\tilde{\eta}} \left[ 
(y-X\Delta-\tilde{\eta})^T \SAR^{-1} (y-X\Delta-\tilde{\eta})+\tilde{\eta}^T\SRW^{-1}\tilde{\eta}
\right]} \\
&=& \min_{\Delta,\tilde{\eta}} \left[ 
\{\tilde{\eta}-( \SAR^{-1}+\SRW^{-1})^{-1}\SAR^{-1}(y-X\Delta)\}^T(\SAR^{-1}+\SRW^{-1})
\{ \tilde{\eta}-( \SAR^{-1}+\SRW^{-1})^{-1}\SAR^{-1}(y-X\Delta)\}
\right.\\
& &\left. +
(y-X\Delta)^T
\left\{\SAR^{-1}-\SAR^{-1}(\SAR^{-1}+\SRW^{-1})^{-1} \SAR^{-1} \right\}
(y-X\Delta)  \right]\\
&=& \min_{\Delta} \left[(y-X\Delta)^T
\left\{\SAR^{-1}-\SAR^{-1}(\SAR^{-1}+\SRW^{-1})^{-1} \SAR^{-1} \right\}
(y-X\Delta)\right].
\end{eqnarray*}

Finally using the Woodbury matrix identity, for symmetric invertible matrices $A$ and $B$, $(A+B)^{-1}=A^{-1}-A^{-1}(A^{-1}+B^{-1})^{-1}A^{-1}$. Thus we have
\[
\left\{\SAR^{-1}-\SAR(\SAR^{-1}+\SRW^{-1})^{-1} \SAR^{-1} \right\}=\left(\SAR+\SRW\right)^{-1}.
\]
The result follows immediately. \hfill $\Box$

\noindent {\bf Proof of Corollary \ref{cor:null}.}


As before write $y$ for $y_{1:n}$ and $X$ for $X_{\tau_{0:d}}$; further let $X_0=X_{\tau_0}$. The value of $\Delta$ that minimises the right-hand side of (\ref{eq:cost_prop}) is
\[
\hat{\Delta}=\{X^T(\SAR+\SRW)^{-1}X\}^{-1}X^T(\SAR+\SRW)^{-1}y.
\]

To further simplify notation let $A=(\SAR+\SRW)^{-1}$ and let $\Phi$ be such that $A=\Phi\Phi^T$ with $\Phi$ invertible; and let $\Psi$  be a matrix such that $\Sigma=\Psi\Psi^T$. Then the reduction in cost over fitting no change is
\begin{eqnarray*}
\lefteqn{C_0-C(\tau_{0:d})=y^T \Big(AX(X^TAX)^{-1}X^TA-AX_0(X_0^TAX_0)^{-1}X_0^TA\Big)y }\\
&=&y^T\Phi^T\Phi^{-T}\Big(AX(X^TAX)^{-1}X^TA-AX_0(X_0^TAX_0)^{-1}X_0^TA\Big)\Phi^{-1}\Phi y = y^T\Phi^T B \Phi y,
\end{eqnarray*}
for the matrix $B=\Phi^{-T}\Big(AX(X^TAX)^{-1}X^TA-AX_0(X_0^TAX_0A)^{-1}X_0^T\Big)\Phi^{-1}$. By standard properties of linear models, as our model includes an intercept term this quadratic form is invariant to adding a constant to all entries of $y$. Thus as our model assumes no change we can, without loss of generality assume the mean of $y$ is the zero vector.

Now it is straightforward to show that $B^2=B$ and that $B$ has rank $d$. Furthermore as under our assumptions $y$ is Gaussian with variance $\Sigma$, $\Phi y$ has variance $\Phi \Sigma \Phi^T$. From standard results for quadratic forms of Gaussian random variables, see for example Theorem 9.5 of \cite{Muller/Stewart:2006}, the distribution of our quadratic form, $y^T\Phi^T B \Phi y$ is
\[
\sum_{i=1}^{d} \alpha_i Z^2_i,
\]
where $\alpha_i$ are the non-zero eigenvalues of $\Phi^T \Psi^T B \Psi \Phi$, and each $Z^2_i$ are independent $\chi^2_1$ distributed random variables. 

The result follows by first noting that as $B$ is a projection its eigenvalues are 1 or 0. Thus $\alpha_i\leq \alpha^+$, where $\alpha^+$ is the largest eigenvalue of $\Phi^T \Psi^T\Psi \Phi$, which by standard results is also the largest eigenvalue of $\Phi\Phi^T\Psi\Psi^T=(\SAR+\SRW)^{-1}\Sigma$. Thus
\[
\sum_{i=1}^{d} \alpha_i Z^2_i \leq \sum_{i=1}^{d} \alpha^+ Z^2_i = \alpha^+ \sum_{i=1}^{d} Z^2_i,
\]
and the right-hand side has the same distribution as $\alpha^+$ times a $\chi^2_{d}$ random variable. If $\Sigma=\SAR+\SRW$ then we further have that $\alpha_i=1$ and hence the distribution is $\chi^2_d$.

To prove the consistency of $\hat{m}$ we need to show that the probability of 
\[
\mathcal{C}_0-\mathcal{C}(\tau_{1:d})< d \beta
\]
jointly for all $d$ and $\tau_{1:d}$ tends to 1. A standard argument \cite[see the proof of Proposition 3.1 in][]{Zheng/Eckley/Fearnhead:2019}, is to use a union bound: 
\begin{eqnarray*}
\Pr(\hat{m}=0)&\geq& 1- \sum_{d=1}^n \frac{n!}{d!(n-d)!} \Pr\left(\chi^2_d> \frac{d\beta}{\alpha^+}\right) \\
&\geq& 1- \sum_{d=1}^n \frac{n!}{d!(n-d)!} \Pr\left(\chi^2_d> dC\log(n)\right) \\
&\geq&1-\sum_{d=1}^n n^d  \exp\left\{ -d\left(\frac{C\log(n)-\sqrt{2C\log(n)-1} }{2}\right)\right\} \\
&\geq& 1-\sum_{d=1}^n \exp\left\{ -d\left(\frac{(C-2)\log(n)-\sqrt{2C\log(n)-1} }{2}
\right)\right\}
\end{eqnarray*}
with the second inequality using a tail bound for a $\chi^2_d$ random variable \cite[Lemma 1 in][]{Laurent/Massart:2000}. The final expression will tend to 1 as $n\rightarrow\infty$ as $C>2$.
\hfill $\Box$

\noindent {\bf Proof of Proposition \ref{prop:alt}.}

 We use the notations $A = (\SAR+\SRW)^{-1}$, $u_1 = u_{\tau_1}$, and write $c_0 = u_0^T A u_0$, $c_{0,1} = u_0^T A u_1$ and $c_1 = u_1^T A u_1$.

The optimal cost is equal to $y^TAy - (X^TAy)^T(X^TAX)^{-1}X^TAy$.
If $X$ is simply a column of ones, $X = u_0$ then $(X^TAX)^{-1} = \frac{1}{c_0}$, and $X^TAy=u_0^TAy$. 

If $X$ is the concatenation of $u_0$ and $u_1$, $X = (u_0\,\,u_1)$ we can compute 
 $$ X^TAX=
  \left[ {\begin{array}{cc}
    c_{0} & c_{0,1} \\
   c_{0,1} & c_{1} \\
  \end{array} } \right]\quad \hbox{and}\quad (X^TAX)^{-1} =\frac{1}{c_{0}c_{1} - c_{0,1}^2}
 \left[ {\begin{array}{cc}
    c_{1} & -c_{0,1} \\
   -c_{0,1}  & c_{0} \\
  \end{array} } \right]\,.$$
  We also have 
  $$\left[ {\begin{array}{cc}
    U_{0} \\
   U_{1} \\
  \end{array} } \right] = \left[ {\begin{array}{cc}
    u_{0}^TAy \\
   u_{1}^TAy \\
  \end{array} } \right] = X^T Ay\,.$$
  Finally
  $$C(\tau_1)=y^TAy - \frac{1}{c_{0}c_{1} - c_{0,1}^2}\Big( U_0 c_1 U_0 - 2 U_0 c_{0,1} U_1 + U_1c_0 U_1\Big).$$
Hence we can write the reduction in cost for fitting a change as
\begin{eqnarray*}
C_0-C(\tau_1)&=&\frac{1}{c_{0}c_{1} - c_{0,1}^2}\Big( c_1 U_0^2  - 2c_{0,1} U_0  U_1 + c_0 U_1^2\Big) - \frac{1}{c_0}U_0^2\\
&=&\frac{1}{c_0^2c_1-c_0c_{0,1}^2} \Big( c_{0,1}^2U_0^2-2c_{0,1}c_0U_0U_1+c_0^2U_1^2\Big).
\end{eqnarray*}
Simple algebraic rearrangement gives the result in (i).

For part (ii) note that $\sum_{i=1}^n v_i=u_0^Tv$, using the definition of $v$ gives 
\[
u_0^Tv=\frac{1}{\sqrt{c_{1}-c^2_{0,1}/c_0}} \left\{c_{0,1}- \frac{c_{0,1}}{c_0}c_0 \right\}=0.
\]
Similarly
\[
v^T(\SAR+\SRW)v= \frac{1}{c_{1}-c^2_{0,1}/c_0} \left\{ c_{1}-2\frac{c_{0,1}}{c_0}c_{0,1}+\left(\frac{c_{0,1}}{c_0}\right)^2c_0
\right\}=1.
\]

Part (iii) is a standard result on the optimality of the weighted least squares estimator. To show it we can directly solve the constrained optimisation problem of maximising $(u_{1}^T{w})^2$ subject to
$u_0^T{w}=0$ and ${w}^T(\SAR+\SRW){w}=1$. Using Lagrange multipliers we have that for constants $\alpha$ and $\delta$
\[
2(u_{1}^T{w})u_{1}=\alpha u_0 +2\delta(\SAR+\SRW){w}.
\]
Defining $\delta'=(u_{1}^T{w})/\delta$, and $\alpha'=-\alpha/(2\delta)$, we get
\[
{w}= \delta' (\SAR+\SRW)^{-1}u_{1} +\alpha' (\SAR+\SRW)^{-1}u_0.
\]
This means that ${w}$ is a linear combination of the vectors $(\SAR+\SRW)^{-1}u_{1}$ and $(\SAR+\SRW)^{-1}u_0$, with the constants uniquely defined by the constraints. However this is the form that $v$ as defined in part (i) takes, hence part (iii) of the proposition holds.
\hfill $\Box$

\noindent {\bf Proof of Theorem \ref{thm:consistency} }

We will first consider the case where $\phi = 0$. For each $n$ introduce the following sets of segmentations of the data:
\[
\mathcal{A}^{n}_{i,m} = \left\{\tau_{1:m}: \min_{j=1,\ldots,m} | \tau_j-\tau^0_i|>(\log n)^2\right\}; ~ i=1,\ldots,m^0,~ m=1,\ldots,m_{\max};
\]
\[
\mathcal{B}^{n}_{m} = \left\{\tau_{1:m}: \max_{i=1,\ldots,m^0} \left( \min_{j=1,\ldots,m} | \tau_j-\tau^0_i| \right) \leq (\log n)^2 \right\}; ~ m=m^0+1,\ldots,m_{\max}.
\]
Thus $\mathcal{A}^{n}_{i,m}$ is the set of segmentations with $m$ changepoints which do not contain a change within a distance $ (\log n)^2$ of the $i$th actual changepoint; and $\mathcal{B}^n_m$ is the set of segmentations with $m>m^0$ changepoints and that have one changepoint within a distance of $(\log n)^2$ of each true changepoint. If a segmentation is in none of these sets then it must have the correct number of chanepoints, and one changepoint within a distance $(\log n)^2$ of each true change. As there are fixed number of these sets, to prove our result we need to show that $\Pr(\hat{\tau}_{1:\hat{m}}\in \mathcal{A}^n_{i,m})  \rightarrow 0$ for each $i$ and $m$; and $\Pr(\hat{\tau}_{1:\hat{m}}\in \mathcal{B}^n_{m})  \rightarrow 0$ for each $m$.

Let $\mathcal{C}(\tau_{1:m})$ denote the unpenalised cost for the segmentation $\tau_{1:m}$, with, for example, $\mathcal{\C}(\tau_{1:m},\tau^0_{1:m^0})$ the unpenalised cost from the segmentation that has the changepoints in the union of $\tau_{1:m}$ and $\tau^0_{1:m^0}$.
We first show that for any $m=m^0+1,\ldots,m_{\max}$, $\Pr(\hat{\tau}_{1:\hat{m}}\in \mathcal{B}^n_{m})  \rightarrow 0$. To do this consider a $\tau_{1:m}\in \mathcal{B}^n_m$, we will compare the cost of this segmentation with that of the true segmentation. As adding changepoints can only reduce the unpenalised cost we have the difference in penalised costs is
\[
\mathcal{C}(\tau_{1:m})+m\beta-\mathcal{C}(\tau^0_{1:m^0})-m^0\beta\geq (m-m^0)\beta - \left(\mathcal{C}(\tau^0_{1:m^0})- \mathcal{\C}(\tau_{1:m},\tau^0_{1:m^0})\right).
\]
Furthermore, by the same argument used in Corollary \ref{cor:null}, $(\mathcal{C}(\tau^0_{1:m^0})- \mathcal{\C}(\tau_{1:m},\tau^0_{1:m^0}))/\alpha$ is stochastically bounded by a $\chi^2_m$ distribution. 

As there are fewer than $(2(\log n)^2)^{m^0}n^{m-m^0}$ segmentations in $\mathcal{B}^n_m$ we have
\begin{eqnarray*}
\lefteqn{\Pr \left(\min_{\tau_{1:m}\in \mathcal{B}^n_m}  \mathcal{C}(\tau_{1:m})+m\beta < \mathcal{C}(\tau^0_{1:m^0}) + m^0\beta)
\right)} & & \\
& \leq & (2(\log n)^2)^{m^0}n^{m-m^0} \Pr( \chi^2_m > (m-m^0)\beta/\alpha) \\
&=& (2(\log n)^2)^{m^0}n^{m-m^0} \Pr( \chi^2_m > (m-m^0)C\log n).
\end{eqnarray*}
By a similar argument to that used in the proof of Corollary \ref{cor:null}, this probability tends to 0 as required.

Now we consider $\tau_{1:m}\in \mathcal{A}_{i,m}^n$. Again we will compare the cost of such a segmentation with that of the true segmentation. Let $\tau^0_{-i}$ denote the set of true changepoints excluding $\tau^0_i$.
\begin{eqnarray*}
\lefteqn{\mathcal{C}(\tau_{1:m})+m\beta-\mathcal{C}(\tau^0_{1:m^0})-m^0\beta \geq \mathcal{C}(\tau_{1:m},\tau^0_{-i})-\mathcal{C}(\tau^0_{1:m^0})+(m-m^0)\beta } & & \\
&=& \{ \mathcal{C}(\tau_{1:m},\tau^0_{-i}) - \mathcal{C}(\tau_{1:m},\tau^0_{1:m^0}) -m^0 \beta\} +
\{ \mathcal{C}(\tau_{1:m},\tau^0_{1:m^0}) - \mathcal{C}(\tau^0_{1:m^0}) + m \beta \}
\end{eqnarray*}

There are fewer than $n^m$ segmentations in $\mathcal{A}_{i,m}^n$, and $(\mathcal{C}(\tau_{1:m},\tau^0_{1:m^0}) - \mathcal{C}(\tau^0_{1:m^0}) )/\alpha$ is stochastically bounded by a $\chi^2_m$ random variable. Thus by the same argument as above we have that 
\[
\Pr\left( \min_{\tau_{1:m} \in \mathcal{A}_{i,m}^n} \mathcal{C}(\tau_{1:m},\tau^0_{1:m^0}) - \mathcal{C}(\tau^0_{1:m^0}) + m \beta < 0 \right) \rightarrow 0.
\]
To show $\Pr(\hat{\tau}_{1:m} \in \mathcal{A}^n_{i,m})\rightarrow 0$ we only need to show
\[
\Pr\left( \min_{\tau_{1:m}\in \mathcal{A}_{i,m}^n} \mathcal{C}(\tau_{1:m},\tau^0_{-i}) - \mathcal{C}(\tau_{1:m},\tau^0_{1:m^0}) -m^0 \beta <0 \right) \rightarrow 0.
\]
By the same argument as used in Proposition \ref{prop:alt}(i), $\mathcal{C}(\tau_{1:m},\tau^0_{-i}) - \mathcal{C}(\tau_{1:m},\tau^0_{1:m^0})=(v^Ty_{1:n})$ for some vector $v=v_{1:n}$. By standard properties of linear models, it is straightforward to show that $v$ has the following properties: (i) $v^T\Sigma^*_nv=1$, where $\Sigma^*_n=\SRW+\SAR$ is the variance of the noise in the fitted model; (ii) $v$ is orthogonal to the column-space of the $X$ matrix for the linear model (\ref{eq:model}) corresponding to the changepoints $\tau_{1:m},\tau^0_{1:m^0}$; (iii) among vectors $v$ that satisfy (i) and (ii) it is the one that maximises the signal for a change at $\tau_i$, i.e. that maximises $(\sum_{t=1}^{\tau_i} v_i)^2$.

If we define $\nu=(\sum_{t=1}^{\tau_i} v_i)^2$, we can bound $\nu$ by choosing any vector $w=w_{1:n}$ that satisfies (ii) and then, after normalising using (i), property (iii) gives $\nu\geq (\sum_{t=1}^{\tau_i} w_i)^2/(w^T\Sigma^*_nw)$.
Let $h=\lfloor (\log n)^2\rfloor$.
We choose such a $w$ defined as $w_j=1$ for $j=\tau_i-h+1,\ldots,\tau_i$, $w_j=-1$ for $\tau_i+1,\ldots,\tau_i+h$, and $w_j=0$ otherwise. The column space of the $X$ matrix in property (ii) contains vectors whose $j$th entries are either identically 0 or identically 1 for  for $j=\tau_i-h+1,\ldots,\tau_i+h$, and hence this vector satisfies property (ii).

Now using the fact that we run DeCAFS with $\phi=0$ and so $\SAR$ is the identity: $w^T\Sigma^*_nw=w^T\SAR w+w^T\SRW w\leq2hc_{\nu}+h^3c_{\eta}/n$, and $\nu\geq h^2/(2hc_{\nu}+h^3c_{\eta}/n)$. Thus there exists $c_1>0$ such that for large enough $n$, $v^Ty_{1:n}$ is normally distributed with $|\mbox{E}(v^Ty_{1:n})|\geq c_1\log n$ and $\mbox{Var}(v^Ty_{1:n})\leq \alpha$. So, for large enough $n$,
\begin{eqnarray*}
\lefteqn{\Pr\left( \min_{\tau_{1:m}\in \mathcal{A}_{i,m}^n} \mathcal{C}(\tau_{1:m},\tau^0_i) - \mathcal{C}(\tau_{1:m},\tau^0_{1:m^0}) -m^0 \beta <0 \right) } & & \\
&\leq & n^m \Pr\left( Z < \frac{1}{\sqrt{\alpha}}\{\sqrt{C\alpha\log n m^0}- c_1 \log n \} \right),
\end{eqnarray*}
where $Z$ is a standard normal random variable. Using standard tail bounds we get that this probability tends to 0 as $n\rightarrow \infty$ as required.

The argument for the case where $\phi>0$ is similar. The differences are just in the definition of the sets $\mathcal{A}_{i,m}^n$ and  $\mathcal{B}^n_m$ which are now
\[
\mathcal{A}^{n}_{i,m} = \left\{\tau_{1:m}: \min_{j=1,\ldots,m} | \tau_j-\tau^0_i|>0 \right\}; ~
\mathcal{B}^{n}_{m} = \left\{\tau_{1:m}: \max_{i=1,\ldots,m^0} \left( \min_{j=1,\ldots,m} | \tau_j-\tau^0_i| \right) =0  \right\}; 
\]
and the final part of the argument that shows 
\begin{equation} \label{eq:probto0}
\Pr\left( \min_{\tau_{1:m}\in \mathcal{A}_{i,m}^n} \mathcal{C}(\tau_{1:m},\tau^0_i) - \mathcal{C}(\tau_{1:m},\tau^0_{1:m^0}) -m^0 \beta <0 \right) \rightarrow 0.
\end{equation}
For this last part we use a different vector $w$ to bound the distribution of $\mathcal{C}(\tau_{1:m},\tau^0_i) - \mathcal{C}(\tau_{1:m},\tau^0_{1:m^0})=(v^Ty)^2$. Our choice of $w$ has $w_{\tau_i}=1$, $w_{\tau_i+1}=-1$ and $w_j=0$ otherwise. We then have $w^T\Sigma^*_n w= w^T\SAR w+w^T\SRW w=2(1-\phi)c_\nu(1-\phi^2)+c_{\eta}/n$. Now as $\phi=\exp\{-c_\phi/n\}\geq 1-c_\phi/n$ we have $w^T\Sigma^*_n w\leq c_1/n$ for some constant $c_1$. Thus $\nu\geq n/c_1$. As this is $O(n)$ it is straightforward to use the same tail bounds of a normal random variable to show (\ref{eq:probto0})

\noindent
{\bf Proof of Proposition \ref{prop:C3} }

If we fix $n$, and let $\Sigma^0$ be the covariance matrix of the generated data then in case (i), $[\Sigma^0]_{ij}=\mbox{Cov}(\zeta(i/n),\zeta(j,n))=c^0_\eta \min{i,j}/n$ if $i\neq j$ and $[\Sigma_0]_{ii}=\mbox{Var}(\zeta(i/n))=c^0_\eta i + c^0_\nu$. Whilst in case (ii), 
\[
[\Sigma^0]_{ij}=\mbox{Cov}(\zeta(i/n),\zeta(j,n))= c^0_\eta \min{i,j}/n
+c^0_\nu (\exp\{-c_\phi^0/n\})^{|i-j|}.
\]
In both cases we can write $\Sigma^0=\SAR^0+\SRW^0$ where $\SAR^0$ is the covariance matrix of an AR(1) process with auto-correlation parameter, $\phi^0=\exp\{-c^0_\phi/n\}$, and marginal variance $c_\nu^0$ and $\SRW^0$ is the covariance matrix of a random walk process with variance parameter $c_\eta^0/n$.
 
We proceed by calculating a bound for the maximum eigenvalue of $\Sigma^{-1}\Sigma^0$, where $\Sigma=\SAR+\SRW$ and $\Sigma^0=\SAR^0+\SRW^0$ are respectively the covariance assumed by DeCAFS and the covariance of the data. We then further bound this as we vary $n$ for the given parameter regimes for the two covariance matrices. We do this first for case (i) where $\phi=\phi^0=0$, then for the case where both autocorrelation parameters are non-zero.

Standard manipulations give that the maximum eigenvalues of $\Sigma^{-1}\Sigma^0$ is also the maximum eigenvalue of $\Sigma^{-1/2}\Sigma^0\Sigma^{-1/2}$, where $\Sigma^{-1/2}$ is a symmetric square root of $\Sigma^{-1}$. If $v$ is an eigenvector of $\Sigma^{-1/2}\Sigma^0\Sigma^{-1/2}$ with eigenvalue $\rho$, then
\[
v^T\Sigma^{-1/2}\Sigma^0\Sigma^{-1/2}v=\rho v^Tv.
\]
Writing $w=\Sigma^{-1/2}v$, we have
\[
\frac{w^T\Sigma^0w}{w^T\Sigma w}=\rho, 
\]
from which we have that we can bound the maximum eigenvalue by
\begin{eqnarray}
\max_{w:|w|=1} \frac{w^T\Sigma^0w}{w^T\Sigma w}& = &
\max_{w:|w|=1} \frac{w^T\SAR^0w+w^T\SRW^0 w}{w^T\SAR w+w^T\SRW w} \nonumber \\
&\leq & \max\left\{\max_{w:|w|=1} \frac{w^T\SAR^0w}{w^T\SAR w}, \max_{w:|w|=1} \frac{w^T\SRW^0 w}{w^T\SRW w} \right\}. \label{eq:max1}
\end{eqnarray}
The first part of the Proposition follows by noting that $\SRW^0=(c^0_\eta/c_\eta)\SRW$, and, if $\phi=\phi^0=0$, $\SAR^0=(c^0_\nu/c_\nu)\SAR$. Hence, 
\[
\max_{w:|w|=1} \frac{w^T\SAR^0w}{w^T\SAR w}=\frac{c_\nu^0}{c_\nu},~~
\max_{w:|w|=1} \frac{w^T\SRW^0w}{w^T\SRW w}=\frac{c_\eta^0}{c_\eta}.
\]

For the case where $\phi^0\neq0$ and $\phi\neq0$ we use a similar argument but, in addition, need to bound $
\max_{w:|w|=1} {w^T\SAR^0w}/{w^T\SAR w}$. Now by similar arguments to above, we have that this is just the largest eigenvalue of $\SAR^{-1/2}\SAR^0\SAR^{-1/2}$, which in turn is
\[
\max_{w: |w|=1} \frac{w^T\SAR^{-1}w}{w^T(\SAR^0)^{-1}w}.
\]
To simplify notation and exposition, fix $n$ and let $r=\phi^0$. Then
\[
\SAR^{-1}=\frac{1}{c_\nu(1-\exp\{-2c_\phi/n\})}K_\phi,\mbox{ and}
(\SAR^0)^{-1}=\frac{1}{c^0_\nu(1-\exp\{-2c^0_\phi/n\})}K_r,
\]
where $K_\phi$ is an $n\times n$ matrix with entries
\[
[K_\phi]_{ij}=\left\{
\begin{array}{cl}
1 & \mbox{if $i=j=1$ or $n$}, \\
1+\phi^2 & \mbox{if $i=j\neq1$ or $n$},\\
-\phi & \mbox{if $|i-j|=1$},\\
0 & \mbox{otherwise,}
\end{array}
\right. 
\]
and similarly for $K_r$. Clearly we have
\begin{equation} \label{eq:max2}
\max_{w: |w|=1} \frac{w^T\SAR^{-1}w}{w^T(\SAR^0)^{-1}w} =
\frac{c^0_\nu(1-\exp\{-c_\phi^0/n\})}{c_\nu(1-\exp\{-c_\phi/n\})} 
\max_{w: |w|=1}
\frac{w^TK_\phi w}{w^TK_r w}.
\end{equation}

Let $v^{(i)}$, for $i=1,\ldots,n$ be the eigenvectors of $K_r$.  Standard results, (see, e.g.,  "Spectral decomposition of Kac-Murdock-Szego Matrices", a technical report by William F Trench available at \url{https://works.bepress.com/william_trench/133/}), are that the eigenvalues are of the form $1-2r\cos \theta_i+r^2$, for some angles $\theta_1,\ldots,\theta_n$.  Furthermore the entries of $v^{(i)}$ satisfy
\[
v^{(i)}_{j-1}- 2\cos \theta_i v^{(i)}_j+v^{(i)}_{j+1}=0,~~ \mbox{for $j=2,\ldots,n$},
\]
with $(2\cos\theta_i-r)v^{(i)}_1=v^{(i)}_2$ and $(2\cos\theta_i-r)v^{(i)}_n= v^{(i)}_{n-1}$.

Straightforward calculations then give
\[
K_\phi v^{(i)} = (1-2\phi\cos \theta_i +  \phi^2)v^{(i)}+\phi(r-\phi)(v^{(i)}_1 e_1 + v^{(i)}_n e_n),
\]
where $e_1$ and $e_n$ are the $n$-vectors of 0s with a 1 in, respectively, the first and $n$th entries.

Now writing $w=\sum_{i=1}^n d_i v^{(i)}$, we have
\[
\frac{w^TK_\phi w}{w^TK_r w}=
\frac{ \sum_{i=1}^n d_i^2 (1-2\phi\cos \theta_i+\phi^2) +\phi(r-\phi) (w_1^2+w_n^2)}
{\sum_{i=1}^n d_i^2 (1-2r\cos \theta_i+r^2)}.
\]
For any $w$ with $|w|=1$ we trivially have that
\[
\frac{ \sum_{i=1}^n d_i^2 (1-2\phi\cos \theta_i+\phi^2)}{\sum_{i=1}^n d_i^2 (1-2r\cos \theta_i+r^2)}\leq \max_\theta\frac{(1-2\phi\cos \theta+\phi^2)}{(1-2r\cos \theta+r^2)}=\max\left\{
\frac{(1-\phi)^2}{(1-r)^2},\frac{(1+\phi)^2}{(1+r)^2}\right\}.
\]
Now if we write $\rho_i=(1-2r\cos \theta_i+r^2)$ for the $i$th eigenvalue of $K_r$, then
\[
\max_{w: |w|=1} \frac{w_1^2}{\sum_{i=1}^n d_i^2 \rho_i}=
\max_{d: |d|=1} \frac{\left(\sum_{i=1}^n d_i v^{(i)}_1\right)^2}{\sum_{i=1}^n d_i^2 \rho_i}=
\left(\sum_{i=1}^n (v^{(i)}_1)^2/\rho_i \right),
\]
where we have first rewritten $w$ and $w_1$ in terms of its expansion in the basis of the eigenvectors of $K_r$, and then used the fact that the maximum is achieved with $d_i \propto v^{(i)}_1/\rho_i$. Using the fact that each $v^{(i)}$ is an eigenvector of $K_r^{-1}$ with eigenvalue $1/\rho_i$,
\[
\left(\sum_{i=1}^n (v^{(i)}_1)^2/\rho_i \right)=[K_r^{-1}]_{11}=\frac{1}{1-r^2}.
\]
By a similar argument for the term involving $w_n^2$ we have
\[
\max_{w: |w|=1} \frac{w^TK_\phi w}{w^TK_r w}
\leq 
\max\left\{
\frac{(1-\phi)^2}{(1-r)^2},\frac{(1+\phi)^2}{(1+r)^2}\right\}+2\max\left\{\phi\frac{(r-\phi)}{1-r^2},0\right\}.
\]
Now using $\phi=\exp\{-c_\phi/n\}$ and $r=\exp\{-c_\phi^0/n\}$ we have
this bound is $(c_\phi/c_\phi^0)^2+(c_\phi-c_\phi^0)/c_\phi^0+O(1/n)$ if $c_\phi>c_\phi^0$ and $1+O(1/n)$ if $c_\phi\leq c_\phi^0$. The result follows trivially by combining this with (\ref{eq:max1}) and (\ref{eq:max2}).

\section{Additional Empirical Results}

\subsection{Parameter Estimation}\label{App:paramEstim}

\begin{figure}[!ht]
    \centering
    \includegraphics[width=13cm]{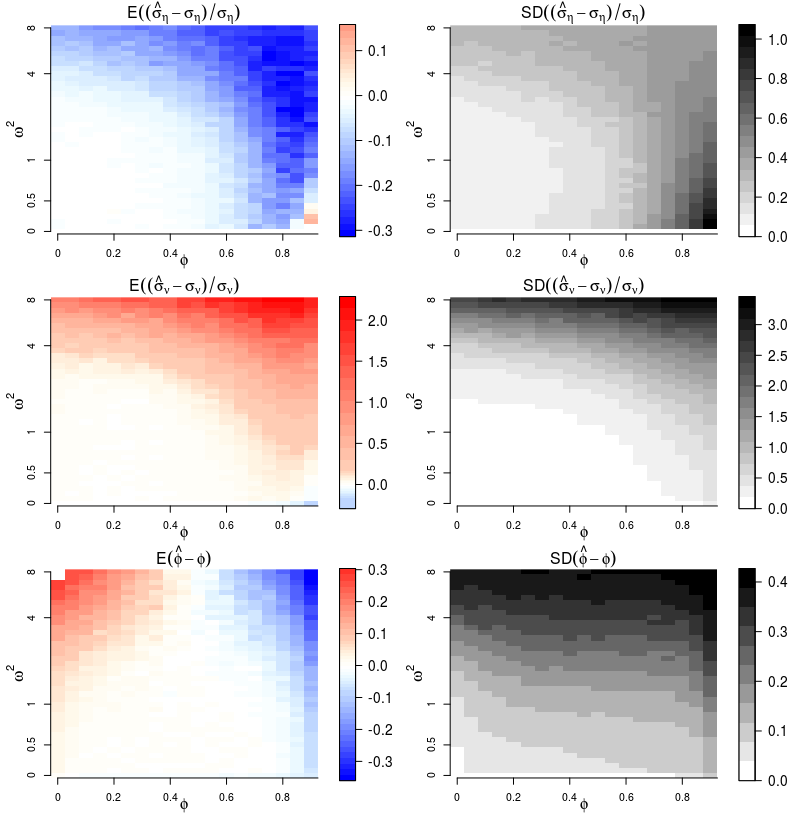}
    \caption{For each cell $1200$ time-series of length $5000$ have been generated under our model (\ref{eq:RWARmodel}) -- (\ref{eq:ARmodel}) with no change. The left column corresponds to an accuracy measure: a percent error for the variances and the bias for the $\phi$ parameter. The right column shows the precision (standard deviation). We chose $K=10$ for the estimators described in Section~\ref{sec:parameter-estimation}.}
    \label{fig:simuParameters}
\end{figure}

We provide a simple simulation study to highlight the behavior of our estimators described in Section \ref{sec:parameter-estimation} for parameters $\sigma_{\eta}$, $\sigma_{\nu}$ and $\phi$. With $K=10$, no change along the data, we simulate $1200$ time-series of length $5000$ for each couple $(\phi,\omega^2)$ on a grid for $\phi \in \{\frac{i-1}{20}\,,\,i=1,...,20\}$ and $\omega^2 = \sigma_{\eta}^2/\sigma_{\nu}^2 \in [0,8]$ with a log scale of $40$ elements. In Figure \ref{fig:simuParameters} we see that as $\omega^2$ and $\phi$ increase, $\sigma_{\eta}$ tends to be underestimated while $\sigma_{\nu}$ overestimated. The $\phi$ parameter is better estimated for small values of $\omega^2$ and intermediate values of $\phi$. The random walk variance is less biased than the AR(1) variance with also a better precision. Notice also that the observed standard deviation for $\phi$ is often greater than $0.1$ and an important deviation to the true parameter of order $0.1-0.2$ is not uncommon.

\begin{figure}
    \centering
    \includegraphics[width=.85\linewidth]{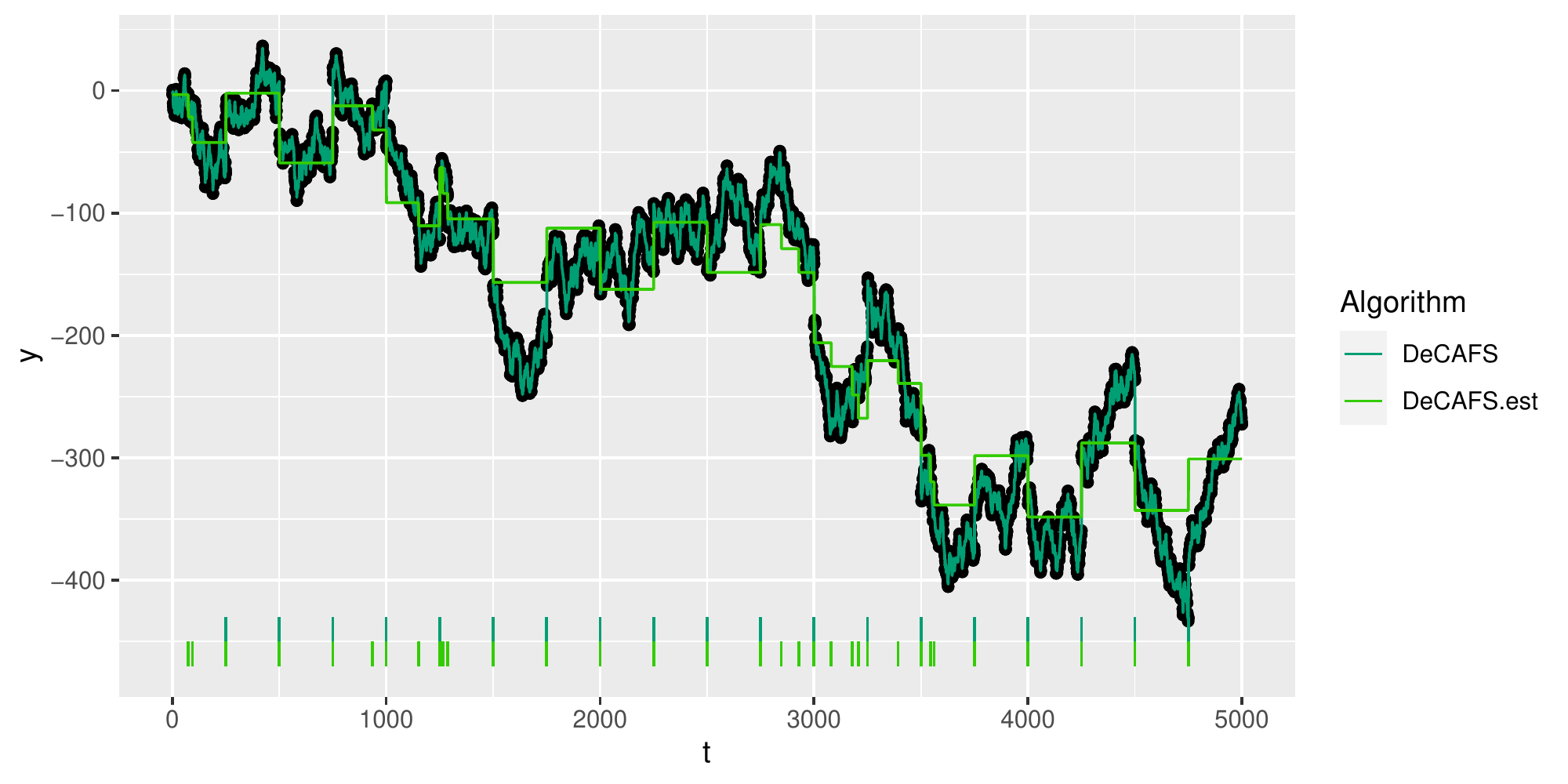}
    \caption{An example of a sequence generated with $\sigma_\eta = 4, \ \sigma_\nu = 2, \ \phi = 0.14$, with relative signal and changepoints estimates of DeCAFS with real parameter values compared to DeCAFS with estimated ones. On this particular sequence, our estimator returns values for initial parameters of $\hat{\sigma}_\eta = 0, \ \hat{\sigma}_\nu = 4.6, \ \hat{\phi} = 0.98$, resulting in a distorted signal estimation.}
    \label{fig:bad-example}
\end{figure}

To see what might happen in case of a distorted parameter estimation, as mentioned in the simulation study of Section \ref{sec:Simulation-Study}, please refer to Figure \ref{fig:bad-example}. We can see there, how even when misspecifying the model, in this case via fitting a pure AR(1) when there was some drift in the signal, we find a distorted signal $\mu$ estimation, however we are still able to reconstruct the changepoint locations relatively well.

\subsection{Additional well-log data segmentation}

\begin{figure}[!t]
    \centering
    \includegraphics[width = .8\linewidth]{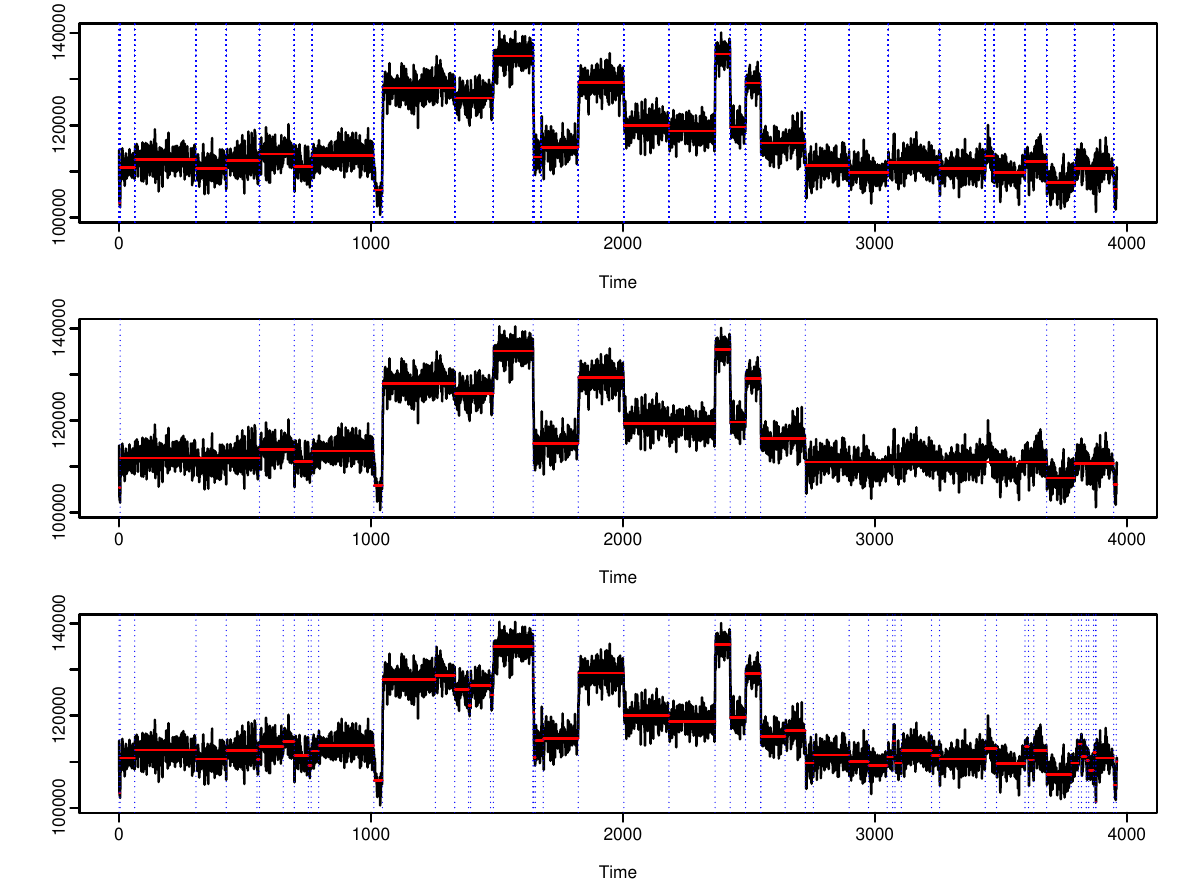}
    \caption{Segmentations of well-log data: Optimal segmentation under square error loss with the default, BIC, penalty (top); segmentation with the AR1-seg method of \cite{chakar2017robust} that models the data as piecewise constant mean with AR(1) noise (middle); optimal segmentation for constant-mean model with WBS2 and the number of changes detected by the steepest drop to low levels criteria of \cite{fryzlewicz2018detecting} (bottom). Each plot shows the data (black line) the estimated mean (red line) and changepoint location (vertical blue dashed lines).}
    \label{fig:well-log-supp}
\end{figure}

In Figure \ref{fig:well-log-supp} we report some additional segmentations of the log-well data described in  Section \ref{sec:Introduction}.

\end{document}